\documentclass[11pt,letterpaper]{article}
\pdfoutput=1
\usepackage{amscd}
\usepackage{amsmath, bbm} 
\usepackage{gensymb}
\usepackage{slashed}
\usepackage{bm}

\usepackage{jheppub} % for details on the use of the package, please

\newcommand{\be}{\begin{equation}}
\newcommand{\ee}{\end{equation}}
\newcommand{\bea}{\begin{eqnarray}}
\newcommand{\eea}{\end{eqnarray}}

\def\beq#1\eeq{\begin{align}#1\end{align}}

\newcommand{\ZP}{Z^\prime}

\begin{document}

\title{
Anomalous $\bm{Z^\prime}$ bosons for anomalous $\bm{B}$ decays
}

\author{Joe Davighi}
\affiliation{DAMTP, University of Cambridge, Wilberforce Road, Cambridge, 
CB3 0WA, UK}
\emailAdd{jed60@cam.ac.uk}

%%%%%%%%%%%%%%%%%%%%%%%%%%%%%%%%%%%%%%%%%%%%%%%%%%%%%%%%%%%%%%%%%%%%%%
\abstract{
Motivated by the intriguing discrepancies in $b\to s \ell\ell$ transitions, the fermion mass problem, and a desire to preserve the accidental symmetries of the Standard Model (SM), we extend the SM by an anomalous $U(1)_X$ gauge symmetry where $X=Y_3+a(L_\mu-L_\tau)/6$. The heavy $Z^\prime$ boson associated with spontaneously breaking $U(1)_X$ at the TeV scale mediates the $b\to s\ell\ell$ anomalies via $\mathcal{O}^\mu_9 \sim\frac{1}{\Lambda^2}(\bar{s}\gamma_\rho P_L b)(\bar{\mu} \gamma^\rho \mu)$.
We show that this model, which features mixed gauge anomalies involving $U(1)_X$ and hypercharge, can be made anomaly-free for any $a\in \mathbb{Z}$ by integrating in a pair of charged fermions whose masses naturally reside somewhere between 1 and 30 TeV.
The gauge symmetry permits only the third family Yukawas at the renormalisable level, and so the light quark masses and mixings are controlled by accidental $U(2)^3$ flavour symmetries which we assume are minimally broken alongside $U(1)_X$.
The lepton sector is not governed by $U(2)$ symmetries, but rather one expects a nearly diagonal charged lepton Yukawa with $m_{e,\mu} \ll m_\tau$. The model does not explain the hierarchy $m_e\ll m_\mu$, but it does possess high quality lepton flavour symmetries that are robust to the heavy physics responsible for generating $m_{e,\mu}$. We establish the viability of these models by checking agreement with the most important experimental constraints. We comment on how the model could also explain neutrino masses and the muon $g-2$.
} 
%%%%%%%%%%%%%%%%%%%%%%%%%%%%%%%%%%%%%%%%%%%%%%%%%%%%%%%%%%%%%%%%%%%%%%

%%%%%%%%%%%%%%%%%%%%%%%%%%%%%%%%%%%%%%%%%%%%%%%%%%%%%%%%%%%%%%%%%%%%%%
\maketitle
%%%%%%%%%%%%%%%%%%%%%%%%%%%%%%%%%%%%%%%%%%%%%%%%%%%%%%%%%%%%%%%%%%%%%%

\normalsize

\section{Introduction}

The requirement of anomaly cancellation imposes stringent constraints on the chiral fermion content of a quantum field theory.  When building models of new physics that extend the Standard Model (SM) gauge symmetry, it seems pragmatic to take our cue from the SM itself and require anomaly cancellation. For family non-universal $U(1)$ extensions of the SM gauge group, the space of anomaly-free charge assignments for the SM fermions has recently been studied both numerically and analytically~\cite{Allanach:2018vjg,Allanach:2019uuu}, and even parametrized explicitly in the special case where three SM singlet fermions are included~\cite{Allanach:2020zna}. Restricting the charges to lie on this anomaly-free subspace often leads to interesting correlations within the phenomenology.

Nonetheless, it is not strictly necessary for the spectrum of SM fermions to be anomaly-free, if the $U(1)$ extension is interpreted as an effective field theory (EFT). By a more familiar analogy, for energies between the masses of the bottom and top quarks one can describe physics using an EFT of the lightest five quarks coupled to the SM gauge group, despite an intricate pattern of gauge anomalies. Integrating out the top quark of course gives rise to operators in this EFT that restore gauge invariance via a four-dimensional version of the Green--Schwarz mechanism~\cite{Green:1984sg,Preskill:1990fr}. In this spirit, we here entertain the possibility that particle physicists might discover a new heavy $Z^\prime$ gauge boson before they discover every fundamental chiral fermion, and so that $Z^\prime$ might have anomalous couplings to the chiral fermions that we know of so far. Lest we forget, the $W$ and $Z$ bosons were discovered at LEP before the Tevatron found the top quark, and the top quark is needed for the SM's anomaly freedom. 

A frequent argument contrary to this scenario is that it is difficult to hide new chiral fermions at suitably high masses, without prematurely breaking electroweak symmetry. (Returning to our 5-flavour SM analogy, the top quark could not have been much heavier than it is, being chiral with respect to $SU(2)_L \times U(1)_Y$.) While it is in general challenging to give heavy enough masses to some complicated set of charged chiral fermions needed for anomaly cancellation, in this paper we will consider a very particular anomalous $U(1)_X$ gauge symmetry that acts as $X=Y_3+a(L_\mu-L_\tau)/6$ on the SM fermions. This choice is phenomenologically motivated `from the bottom-up' (\S \ref{sec:charges}), with our three main motivations being summarized in the next few paragraphs. For this anomalous $U(1)_X$ charge assignment we can restore anomaly cancellation through a pair of chiral fermions whose masses naturally reside at the heavy scale of $U(1)_X$ breaking (\S \ref{sec:UV}). Crucial to this construction is that the various mixed gauge anomalies can be cancelled by BSM fermions that are vector-like (but not neutral) with respect to the SM gauge symmetry, being chiral only under $U(1)_X$.

\paragraph{1. Neutral current $B$-anomalies.}
Our primary phenomenological reason for invoking a TeV scale $Z^\prime$ boson, and for choosing its particular set of anomalous charges, is to explain a collection of recent measurements of rare semi-leptonic $B$-meson decays, all involving $b\to s \ell\ell$ transitions, that are discrepant with the SM predictions. The observables showing discrepancies, which we henceforth refer to as the neutral current $B$-anomalies (NCBAs), include the lepton flavour universality violating (LFUV) ratios $R_K^{(*)} = BR(B\rightarrow K^{(*)} \mu^+ \mu^-) / BR(B\rightarrow K^{(*)}e^+ e^-)$~\cite{Aaij:2017vbb,Aaij:2019wad,Aaij:2021vac}, as well as the $B$-meson branching ratios $BR(B_s \rightarrow \mu^+ \mu^-)$~\cite{Aaboud:2018mst,Chatrchyan:2013bka,CMS:2014xfa,Aaij:2017vad,LHCbtalk}, $BR(B_s \rightarrow \phi \mu^+ \mu^-)$~\cite{Aaij:2015esa,CDF:2012qwd}, and angular distribution observables for $B\rightarrow K^{(*)} \mu^+ \mu^-$ decays~\cite{Aaij:2013qta,Aaij:2015oid,Aaboud:2018krd,Sirunyan:2017dhj,Khachatryan:2015isa,Bobeth:2017vxj}. The recent LHCb measurement of $R_K$ alone sits $3.1\sigma$ lower than its SM prediction of approximately 1, for which the theoretical uncertainties are negligible. A highly conservative method for combining the various $b\to s \ell\ell$ anomalies, including the look-elsewhere effect, estimates their global significance to be 3.9$\sigma$~\cite{Lancierini:2021sdf}. 

\sloppy The combined set of discrepancies are by and large self-consistent with a common new physics origin involving muons. In particular, including a new physics contribution for either $C_L$ or $C_9$, which are the Wilson coefficients of $\mathcal{O}_L \sim \frac{1}{\Lambda^2}(\bar{s}\gamma_\rho P_L b)(\bar{\mu} \gamma^\rho P_L \mu)$ and $\mathcal{O}_9 \sim\frac{1}{\Lambda^2}(\bar{s}\gamma_\rho P_L b)(\bar{\mu} \gamma^\rho \mu)$ respectively, fits the observables $R_K$, $R_K^{\ast}$, and $BR(B_s \rightarrow \mu^+ \mu^-)$, for which the theoretical uncertainties are smallest, better than the SM by a pull of 4.7$\sigma$ (for $C_L$) or 4.1$\sigma$ (for $C_9$)~\cite{Cornella:2021sby,Altmannshofer:2021qrr}. If the other discrepant observables are included, the pull is far greater still, and is found to be essentially the same for new physics in $C_L$ or $C_9$~\cite{Altmannshofer:2021qrr}.\footnote{A complimentary way to assess the statistical signifance of the NCBAs is to estimate a $p$-value for the likelihood of the SM, including a suitably large set of non-discrepant measurements in the fit also. One recent estimate~\cite{Allanach:2021kzj}, which included the full set of $B$-decay data as well as hundreds of observables showing good agreement with the SM, found a $p$-value of 0.00057 for the SM, suggesting a tension at the 3.4$\sigma$ level.} 

Either $C_L$ or $C_9$ (or any similar combination of Wilson coefficients) can be obtained by integrating out a heavy $\ZP$ boson that couples to muons and to a flavour-changing $b\bar{s}$ quark current, where the heavy gauge boson arises from spontaneously breaking some $U(1)_X$ gauge symmetry at the scale of a few TeV. This option, along with scalar and vector leptoquark mediators, has been thoroughly explored in recent model-building literature. The vast majority of these models (beyond the early simplified models) have focussed on gauging anomaly-free $U(1)_X$ symmetry groups. In this paper, we initiate a complimentary study of anomalous $\ZP$ models that can be made anomaly-free by integrating in chiral fermions at the TeV scale of the $\ZP$ boson.

\paragraph{2. Third family alignment and the SM Yukawas.}
It is easy to explain the NCBAs alone using a $Z^\prime$ boson; one just needs the requisite couplings to muons and to $b\bar{s}$, with the latter coupling being small enough to circumvent the constraint from $B_s-\overline{B_s}$ mixing. To go further, there are more qualitative general features in the NCBA data, viewed in the context of precision flavour data as a whole, that we wish to incoporate into our $\ZP$ model.

One striking feature is that all the discrepant observables involve third family quark transitions, and can be explained without any new physics coupling to the first family. Together with the strengthening signs of LFUV, this means that the new physics responsible for the NCBAs breaks the $U(3)^5$ family symmetries of the SM gauge sector in a rather specific way, hinting at the existence of a new family-dependent force that couples strongly to the third family and weakly to the first family. It is tempting to suppose that such family-dependent dynamics are linked to the only other source of $U(3)^5$ breaking in the SM, namely the Yukawa couplings of the SM. We here explore such a connection in the context of a $\ZP$ model, first demanding that the $U(1)_X$ gauge symmetry allows renormalisable Yukawa couplings for the third family alone. The Yukawas for the light families  come from higher-dimensional operators which, for the quark sector, will be governed by $U(2)^3$ family symmetries that are well-known to give a good account of the quark mass and mixing data~\cite{Pomarol:1995xc,Barbieri:1995uv,Barbieri:2011ci,Blankenburg:2012nx,Barbieri:2012uh}.

\paragraph{3. Accidental symmetries of the SM.}
A second striking feature of the NCBAs and related data is that, despite the compelling evidence for LFUV, there is no sign whatsoever that any of the SM's accidental global symmetries $U(1)_B \times U(1)_e \times U(1)_\mu \times U(1)_\tau$, namely baryon number and the three individual lepton numbers, are violated. There are very strong bounds~\cite{Zyla:2020zbs} on lepton flavour violation (LFV) from processes like $\mu\to e\gamma$, $\mu \to 3e$, and $\tau\to 3\mu$, which suggest that lepton number symmetries remain intact up to a scale of thousands of TeV or higher. The non-observation of decaying protons~\cite{Zyla:2020zbs} suggests $U(1)_B$ remains intact at far higher energies still.  It is therefore highly desirable for $U(1)_B \times U(1)_e \times U(1)_\mu \times U(1)_\tau$ to remain a high-quality accidental symmetry of any new physics model designed to explain the NCBA data (if we do not wish various operators and mixing angles to be artificially tuned to evade the bounds). 

Achieving this is particularly challenging for leptoquark models, which permit renormalisable operators violating all of these accidental symmetries. One way out is for the leptoquarks themselves to be charged under some family-dependent gauge symmetry~\cite{Davighi:2020qqa,Greljo:2021xmg}. For $\ZP$ models there is no danger of baryon number violation (beyond that expected in the SM), but a generic family-dependent $U(1)_X$ gauge symmetry will not necessarily protect the individual lepton numbers. A host of elegant $\ZP$ solutions that predict LFUV without LFV have been based on gauging linear combinations of the lepton number symmetries themselves, chiefly the anomaly-free combination $L_\mu-L_\tau$~\cite{Baek:2001kca,Ma:2001md,Harigaya:2013twa,Altmannshofer:2014cfa,Altmannshofer:2014pba,Crivellin:2015mga,Altmannshofer:2015mqa,Crivellin:2016ejn,Crivellin:2018qmi,Altmannshofer:2019xda,Altmannshofer:2019zhy}. Such a $\ZP$ can mediate the NCBAs if a fourth-family of vector-like quarks is included to generate the requisite quark flavour violation via mixing.

\subsection*{An anomalous $\bm{\ZP}$ solution}
In this paper we explore one simple way to tie all these observations together via a heavy $\ZP$ boson that arises from spontaneously breaking a $U(1)_X$ gauge symmetry with family-dependent couplings. The model builds on previous attempts to connect the NCBAs with fermion masses using a $U(1)_X$ gauge boson. In~\cite{Allanach:2018lvl,Davighi:2019jwf} a $\ZP$ model was proposed based on gauging the anomaly-free $X=Y_3$ symmetry, where $Y_3$ denotes third family hypercharge quantum numbers, thus allowing renormalisable third family Yukawas.
However, there is an obvious snag in this explanation, which is that the NCBAs involve muons, not tauons. If one insists that $U(1)_X$ is anomaly-free, then the simplest ways to generate a big enough muon coupling while preventing large LFV between $\mu$ and $\tau$ involve sacrificing a renormalisable tauon Yukawa one way or another~\cite{Allanach:2018lvl,Allanach:2019iiy}.\footnote{One option is to introduce a $U(1)_X$ charge for muons, while deforming the third family lepton charges away from hypercharge in order to maintain anomaly-freedom as in~\cite{Allanach:2019iiy}. The other option is for only the third family to remain charged, in which case the unique charge assignment is Third Family Hypercharge~\cite{Allanach:2018lvl}, but to generate the coupling to muons by misaligning the weak eigenbasis from the mass eigenbasis. This of course violates the lepton flavour symmetries that we wish to preserve, and moreover implies the tauon Yukawa is small relative to off-diagonal elements, and so must be fine-tuned since it is allowed at the renormalisable level.}

In this paper, we instead choose to sacrifice anomaly-freedom of the $\ZP$ model at the weak scale. This way, one can attain renormalisable Yukawas for the top, bottom, and tauon (only), as well as a direct $U(1)_X$ coupling to muons needed to explain the NCBAs, without any other SM fermions being charged under $U(1)_X$. The resulting pattern of gauge anomalies requires new chiral fermions to restore $U(1)_X$ gauge invariance at the multi-TeV scale associated to its breaking. Identifying a suitable set of such heavy chiral fermions is now an essential part of model building if we gauge an anomalous $U(1)_X$ symmetry. To simplify the challenge, we require that all the gauge anomalies that are linear in the $U(1)_X$ charges vanish. Together with our requirements of third family Yukawas and charged muons, this in fact leaves only a single family of $U(1)_X$ assignments, which acts on the SM fermions as
$$
X = Y_3 + \frac{a}{6}(L_\mu-L_\tau), \quad a\in \mathbb{Z} \, .
$$
This assignment is a linear combination of two anomaly-free charge assignments $Y_3$ and $L_\mu - L_\tau$, but it is itself anomalous due to non-vanishing quadratic and cubic anomaly coefficients. The model inherits attractive features from models based on either $Y_3$ or $L_\mu-L_\tau$, specifically the third-family alignment and a connection to fermion mass hierarchies from the $Y_3$ component, and a mechanism for preserving lepton flavour from the $L_\mu-L_\tau$ component.

In \S \ref{sec:UV} we show how a pair of electrically charged fermions that are chiral under $U(1)_X$ can always be found to restore anomaly freedom for any integer $a$. Because these fermions are vector-like with respect to the SM gauge symmetry, they can be given masses via Yukawa couplings to a SM singlet scalar. This anomalous $\ZP$ model, in contrast to its anomaly-free counterparts, necessarily features a pair of long-lived charged fermions whose masses are tied to the scale of $U(1)_X$ breaking; in particular, their masses naturally lie between 1 and 30 TeV.\footnote{The lower limit of 1 TeV charged fermions corresponds to Yukawa couplings of $10^{-1}$.
}
 
In \S\S \ref{sec:yuk} and \ref{sec:LFV} we turn to the issue of fermion masses. Beyond the renormalisable third family Yukawa couplings, the light fermion masses and mixings must come from higher-dimensional operators in the TeV scale EFT that derive from a further layer of new physics at a higher scale $\Lambda\approx 100$ TeV. We remain agnostic about the details of this UV physics, limiting our study to the EFT of the SM fields and a small set of spurion fields. These spurions are charged under both the $U(1)_X$ gauge symmetry and the $U(2)^3$ accidental symmetries of the renormalisable quark sector of our model, and so the quark Yukawa textures closely resemble those of other $U(2)^3$-based models.
For the charged leptons, however, our choice of $U(1)_X$ explicitly breaks the corresponding $U(2)^2$ leptonic symmetries in the gauge sector. A spurion analysis reveals that the effective lepton Yukawa matrix is nearly diagonal (up to parts per mille corrections), with the first and second family leptons being hierarchically lighter than the tauon. This means that the $\ZP$ boson has lepton flavour conserving couplings, even though these couplings violate lepton flavour universality. 

Moreover, in \S \ref{sec:correlations} we estimate that lepton flavour violating operators induced by integrating out generic heavy physics at $\Lambda\approx 100$ TeV are extremely suppressed, thanks to the $U(1)_X$ symmetry governing the TeV scale EFT. The model therefore has a sufficiently high quality $U(1)_e \times U(1)_\mu \times U(1)_\tau$ global symmetry consistent with the stringent bounds, a strength that it shares with traditional models based on gauging $L_\mu-L_\tau$. In our case, the quality of these global symmetries is not quite a {\em fait accompli}, as in most $L_\mu-L_\tau$ models, because our simultaneous attempt to shed light on the fermion mass problem necessitates complicated new physics at a relatively low scale around 100 TeV.

We remark that our setup shares qualitative similarities with a string-inspired model~\cite{Celis:2015eqs} that also explains the $b \to s \ell\ell$ anomalies via the couplings of an anomalous $Z^\prime$. However, the $Z^\prime$ in~\cite{Celis:2015eqs} is but one of many extra $U(1)$ gauge fields arising from an underlying dynamics of intersecting D-branes, with a very different set of fermion charges that is determined by the underlying brane geometry in a distinctly `top-down' fashion. The anomalous $Z^\prime$ that we propose in this paper is motivated rather from the bottom-up, in sharp contrast to string-inspired models, by the combination of the NCBAs and the flavour structure and accidental symmetries observed in Nature. Beyond explanations for the NCBAs, we note in passing that anomalous $U(1)$ extensions have been considered as dark matter portals  in~\cite{Ismail:2016tod,Ellis:2017tkh,Ismail:2017ulg}, and the collider constraints on some minimal anomalous $U(1)$ models have been considered in~\cite{Ekstedt:2017tbo,Ismail:2017fgq,Michaels:2020fzj}. 

%%%%%%%%%%%%%%%%%%%%%%%
%%%%%%%%%%%%%%%%%%%%%%%
%%%%%%%%%%%%%%%%%%%%%%%

\section{Gauging an anomalous {\boldmath$Y_3+a(L_\mu-L_\tau)/6$} symmetry} \label{sec:charges}

We begin with the following short wishlist for our family non-universal $U(1)_X$ charge assignment, following the motivations outlined in the Introduction:
\begin{enumerate}
\item We require a non-zero charge for the left-handed second family lepton doublet $\ell_2$,  
\begin{equation}
L_2 \neq 0,
\end{equation}
since a LH component of the muon current is strongly preferred by (muon-only) fits to the NCBA data~\cite{Altmannshofer:2021qrr}. We also allow the charge $E_2$ of the right-handed second family charged lepton $e_2$ to be non-zero. (We use lower case letters to denote the SM fields, and the corresponding capital letters to denote their charges under $U(1)_X$.)
\item We require Yukawas for each third family SM fermion, but no other Yukawa couplings at the renormalisable level,
\be
\mathcal{L}_\text{Yuk} = y_t \, \overline{q_3} h^c u_3 + y_b\, \overline{q_3} h d_3 + y_\tau\, \overline{\ell_3} h e_3  + \text{h.c.} \label{3famYuk}
\ee
This constrains the $U(1)_X$ charges
\begin{align}
Q_3 - U_3 &= -H, \label{eq:t_yuk} \\
Q_3 - D_3 &= H, \label{eq:b_yuk}  \\
L_3 - E_3 &= H, \label{eq:tau_yuk} 
\end{align}
where $H$ is the $U(1)_X$ charge of the SM Higgs field (in a convention where $h$ has positive hypercharge).
\item Of the SM fermions, we only allow non-zero charges for the third family and the second family leptons. Not charging any light quarks brings threefold benefits: (i) FCNCs involving the light quarks remain heavily suppressed as required by {\em e.g.} $\Delta F = 2$ meson mixing measurements; (ii) the quark sector exhibits accidental $U(2)^3$ symmetries at the renormalisable level, whose minimal breaking can give a good account of quark masses and mixings; and (iii) the $\ZP$ couples very weakly to valence quarks and so its production in $pp$ collisions is suppressed, consistent with no resonances being observed in, say, 13 TeV $pp\to \mu\mu$ searches at the LHC~\cite{Aad:2019fac}. By not charging electrons one avoids sizeable couplings of the SM $Z$ boson to electrons that would otherwise be induced by the $Z-Z^\prime$ mixing in our model, which are well-constrained by precision LEP data~\cite{Zyla:2020zbs}.
Together with the Higgs, this leaves us eight $U(1)_X$ charges to determine,
\be
\{Q_3, U_3, D_3, L_2, L_3, E_2, E_3,H\}.
\ee
Note that, since all the other SM fermions are chargeless, point 2. requires 
\be
H \neq 0.
\ee
\item As outlined in the Introduction, we want our model of the NCBAs to have some natural mechanism for suppressing LFV, while accounting for the LFUV observed in $R_{K^{(\ast)}}$. One way to achieve this is to gauge a lepton-flavoured $U(1)_X$ symmetry, under which all three lepton families transform with different charges~\cite{Altmannshofer:2019xda}. For the model we study here the story is a little more subtle, because the electron and muon Yukawas are absent at the renormalisable level, meaning that higher-dimension operators are required with a relatively light suppression scale. But nonetheless, we see in \S \ref{sec:LFV} how the effective charged lepton Yukawa is diagonal to a good approximation. Moreover, in \S \ref{sec:correlations} we show that LFV operators induced by integrating out the light-Yukawa-generating heavy physics are expected to have sufficiently tiny coefficients. Together, these features mean that LFV will be highly suppressed in the model.
\end{enumerate}
Items 1, 2, and 3 on this wishlist are inconsistent with anomaly cancellation, which we must therefore abandon at the electroweak scale. 
Furthermore, it is not enough to only allow the cubic $U(1)_X$ anomaly and the mixed $U(1)_X$-gravitational anomaly to be non-zero, which could be cancelled using SM singlet fermions, as explored {\em e.g.} in Appendix A of~\cite{Altmannshofer:2019xda}.\footnote{This also means that anomaly cancellation in this framework lies beyond the jurisdiction of~\cite{Allanach:2020zna}, in which the SM fermion sector was augmented only by SM singlets.} Rather, there are necessarily mixed anomalies involving SM gauge bosons, as we discuss next. 

For a $U(1)_X$ extension of the SM gauge group, there are six independent local\footnote{There is also a potential global anomaly in the SM gauge symmetry, whose vanishing requires an even number of fields with $SU(2)_L$ isospin quantum numbers $j=2r+1/2$, $r\in \mathbb{Z}$~\cite{Witten:1982fp}. It has recently been shown by computing the spin-bordism groups $\Omega_5^\text{Spin}(B(G_\text{SM}\times U(1)))$ that $U(1)_X$ extensions of the SM suffer from no further global anomalies~\cite{Davighi:2019rcd}, and so, provided the $SU(2)_L$ condition is satisfied by any BSM chiral fermions we add, we only need to worry about local anomaly cancellation.} gauge anomalies involving the $X$ gauge field and the SM gauge fields or gravity. Using the Yukawa conditions (\ref{eq:t_yuk}--\ref{eq:tau_yuk}) to eliminate the variables $U_3$, $D_3$, and $E_3$, we can write the six anomaly coefficients in terms of the remaining five variables. They are as follows,
\begin{align}
&SU(3)^2 \times U(1)_X: \quad &&\mathcal{A}_{SU(3),X} = 0, \label{eq:colour} \\
&SU(2)_L^2 \times U(1)_X: \quad &&\mathcal{A}_{SU(2),X} = 3Q_3+L_2+L_3, \label{eq:su2} \\
&U(1)_Y^2 \times U(1)_X: \quad &&\mathcal{A}_{YYX} = -3(3Q_3-L_2+L_3+2E_2), \label{eq:yyx} \\
&U(1)_Y \times U(1)_X^2: \quad &&\mathcal{A}_{YXX} = -L_2^2+E_2^2-2H(L_3+3Q_3),  \label{eq:quadratic} \\
&U(1)_X^3: \quad &&\mathcal{A}_{XXX} =2L_2^3+L_3^3-E_2^3+3HL_3^2+H^3-3H^2(L_3+6Q_3),  \label{eq:cubic} \\
&\text{Gravity-}U(1)_X: \quad &&\mathcal{A}_{\text{grav},X} = 2L_2+L_3-E_2+H. \label{eq:grav}
\end{align}
Notably, the mixed anomaly involving $SU(3)$ vanishes identically, by virtue of the constraints from the top and bottom Yukawas. This is fortunate, because cancelling a mixed anomaly with $SU(3)$ in the UV would mandate new chiral coloured states. Even for fermion masses of several TeV (appropriate for our heavy states, since they will necessarily be chiral under $U(1)_X$), the bounds on new coloured states are strong (see {\em e.g.} bounds from gluino searches~\cite{Sirunyan:2019ctn,Sirunyan:2019xwh,ATLAS:2019vcq}). 

The remaining five anomalies involving $U(1)_X$ and the electroweak gauge bosons (and gravity) do not vanish. Of course, the more non-vanishing triangle anomalies that we allow, the more difficult becomes the task of cancelling the anomalies at higher energies via chiral fermions that can be given heavy masses -- as is ultimately necessary to ensure consistency of the gauge symmetry. 

In order to progress towards a concrete model that can be made anomaly free at the TeV scale, we make a big simplifying assumption which is that the three anomalies (\ref{eq:su2}, \ref{eq:yyx}, \ref{eq:grav}) that are linear in the $U(1)_X$ charges should vanish at the weak scale. This pragmatic choice eliminates three further degrees of freedom to leave a one-parameter family of solutions for the charges of the SM fields, up to an overall normalisation. Choosing a normalisation in which the quark charges coincide with third family hypercharge ($Y_3$), we have
\begin{align}
&Q_3 =1/6,  &&U_3=2/3,  &&&D_3=-1/3, \\
&L_3 =(-3-a)/6,  &&E_3=(-6-a)/6,  &&&H=1/2 \nonumber \\
&L_2 = a/6, &&E_2=a/6
\end{align}
where $a\in\mathbb{Z}$.
Importantly for the NCBA fit, the second family leptons are required to have vector-like charges.
The charge assignment we have arrived at corresponds to gauging 
\be
X = Y_3 + \frac{a}{6}(L_\mu-L_\tau),
\ee
that is a linear combination of third family hypercharge and $L_\mu-L_\tau$.\footnote{
We note in passing that our anomalous $U(1)_X$ model shares phenomenological similarities with anomaly-free models based on gauging $(B-L)_3+\alpha(L_\mu-L_\tau)$~\cite{Duan:2018akc}, and $(B_1+B_2-2B_3)+\beta (L_\mu-L_\tau)$~\cite{Crivellin:2015lwa}, $\alpha,\, \beta\in \mathbb{Q}$. Both models also couple to $U(2)^3$-preserving quark currents, and to muons via an $L_\mu-L_\tau$ component. 
}

This choice will preserve attractive features of models based on either of these factors. For example, as in third family hypercharge models~\cite{Allanach:2018lvl,Davighi:2019jwf,Allanach:2019iiy} we expect the first two families of each fermion type to be light relative to the weak scale and quark mixing angles to be small, which we explore in more detail in \S \ref{sec:yuk}.\footnote{In the anomaly-free third family hypercharge models, we emphasize that the tauon does not have a renormalisable Yukawa coupling~\cite{Allanach:2019iiy} (or else significant fine-tuning in the lepton Yukawa is needed to explain $b\to s\mu\mu$ anomalies without contravening LFV bounds~\cite{Allanach:2018lvl}).} As in $L_\mu-L_\tau$ models~\cite{Baek:2001kca,Ma:2001md,Harigaya:2013twa,Altmannshofer:2014cfa,Altmannshofer:2014pba,Crivellin:2015mga,Altmannshofer:2015mqa,Crivellin:2016ejn,Crivellin:2018qmi,Altmannshofer:2019xda,Altmannshofer:2019zhy} we obtain muon- and tauon-specific new physics, enabling significant LFUV effects while preventing LFV (\S \ref{sec:LFV}).\footnote{While $(L_\mu-L_\tau)$-based models for explaining the NCBAs must invoke a fourth family of quarks (or something similar) in order to achieve the observed $b\to s$ quark flavour violation~\cite{Altmannshofer:2014cfa}, the model we introduce here already has the requisite quark flavour violation without the extra quarks, because the $Z^\prime$ has flavour non-universal couplings to quarks.} But, unlike gauging $Y_3$ or $L_\mu-L_\tau$ separately, the combination we consider here is not anomaly-free.
Specifically, both the non-linear anomaly coefficients (\ref{eq:quadratic}, \ref{eq:cubic}) receive non-vanishing contributions from the SM fermions,
\be \label{eq:gauge anomalies}
(\mathcal{A}_{YXX})^{\mathrm{SM}} = \frac{a}{6}, \qquad (\mathcal{A}_{XXX})^{\mathrm{SM}} = \frac{a}{4}\, .
\ee
We see that both are proportional to the muon charge $a/6$, meaning the anomalous structure of this $Z^\prime$ model is tightly tied to the NCBA measurements.

\section{Anomaly cancellation at the weak scale} \label{sec:IR}

The fact that we gauge an anomalous $U(1)_X$ symmetry, whose anomaly cancellation via chiral fermions is postponed until the TeV scale (\S \ref{sec:UV}), leaves vestiges in the effective field theory (EFT) that describes the SM fields plus the $Z^\prime$ boson.
Integrating out the heavy chiral fermions will generate terms in the effective action that make gauge invariance manifest in the EFT. The form of such terms in 4d EFTs like the SM was studied in detail by Preskill~\cite{Preskill:1990fr}, while this general procedure for cancelling anomalies using bosonic fields is usually referred to as the Green--Schwarz mechanism in honour of its string theoretic origins~\cite{Green:1984sg}.

For the particular $X=Y_3+a(L_\mu-L_\tau)/6$ anomalous gauge symmetry that we wish to gauge, a suitable pair of Green--Schwarz terms that restore $U(1)_X$-invariance of the EFT are given by
\be \label{eq:GS}
8\pi^2 \mathcal{L}_{\text{GS}} =\frac{a}{6}\, g_X^2 g^\prime \frac{\pi}{v_\Phi}  F_Y\wedge  F_X+ \frac{a}{4}\, g_X^3\frac{\pi}{v_\Phi} F_X \wedge F_X,
\ee
where $\pi(x)$ is a SM singlet pseudoscalar field, 
$F_X = \partial_{[\mu}X_{\nu]} dx^\mu \wedge dx^\nu$ is the field strength for the $U(1)_X$ gauge field and $g_X$ is its gauge coupling, and similarly $F_Y$ denotes the field strength for hypercharge, with  $g^\prime$ denoting the hypercharge gauge coupling as usual. 
The effective lagrangian $\mathcal{L}+\mathcal{L}_{\text{GS}}$ is invariant under $U(1)_X$ gauge transformations 
\begin{align} \label{eq:phi shift}
X_\mu &\mapsto X_\mu + \frac{1}{g_X}\partial_\mu \theta, \\
\pi(x) &\mapsto \pi(x)+ v_\Phi\theta.
\end{align}
The non-linearly realised shift symmetry\footnote{The shift transformation of $\pi$ implies that $\pi(x)/v_\Phi$ is a circle-valued field, since it must be well-defined with respect to the `large gauge-transformation' $\theta \sim \theta+2\pi$.} of $\pi$ signals that $U(1)_X$ must be spontaneously broken, and indeed $\pi$ is nothing but the Goldstone boson associated with $U(1)_X$ breaking at the TeV scale. 
The field $\Phi(x)$ that Higgses $U(1)_X$ can be expanded about its vev as 
\be \label{eq:vev}
\Phi(x)=\frac{1}{\sqrt{2}}(v_\Phi+\sigma(x))e^{i\pi(x)/v_\Phi},
\ee
with the massless mode $\pi$ parametrizing its phase as usual. Of course, the Goldstone $\pi$ may be traded for the longitudinal component of the massive $X$ boson by instead fixing via the unitary gauge. 

The lagrangian is, however, not yet gauge invariant. The mixed anomaly means that under $U(1)_Y$ transformations the effective action shifts by a term proportional to $F_X \wedge F_X$. Now, at intermediate energy scales $v < E < v_\Phi$, hypercharge should be linearly-realised meaning that there is no Goldstone mode and so one cannot restore $U(1)_Y$-invariance by adding a Green--Schwarz term like those in (\ref{eq:GS}). This apparent puzzle has a well-known solution, which is that there exists a local counterterm that can be added to the effective action which restores hypercharge invariance~\cite{Preskill:1990fr},
\be \label{eq:CS}
8\pi^2\mathcal{L}_{\text{c.t.}} =\frac{a}{6}\, g_X^2 g' A_Y \wedge A_X \wedge F_X,
\ee
where $A_Y$ and $A_X$ denote the gauge fields themselves for hypercharge and the $U(1)_X$ symmetry respectively. Note that this term, unlike the pair of dimension-5 terms above, is dimension-4, and so is not suppressed by the heavy mass scale. Its coefficient cannot, after all, become mass-suppressed at low energies, because the term is needed to maintain $U(1)_Y$ invariance all the way down to the electroweak scale.

The Green--Schwarz terms (\ref{eq:GS}, \ref{eq:CS}), which manifest the anomalous structure of our SM extension at low-energies, have {\em a priori} important phenomenological effects that distinguish this model from anomaly-free $Z^\prime$ models. Most notably, they give rise to triangle diagrams replicating each non-vanishing gauge anomaly, which, upon electroweak symmetry breaking and diagonalisation of the neutral gauge boson mass terms, give rise to $Z^\prime \to Z \gamma$ and $Z^\prime \to ZZ$ decays. We comment on these decays in \S \ref{sec:decay}, but the branching ratios will be so small that these processes, while distinctive, do not constitute phenomenologically relevant decay modes of the $Z^\prime$. 

A second hallmark of the fact that $U(1)_X$ is anomalous is that the heavy scale $v_\Phi$ at which the anomalous gauge boson resides cannot be taken arbitrarily high, as can be done consistently for an anomaly-{\em free} spontaneously broken $U(1)_X$ symmetry (simply by taking the vev to be as high as desired). This is ultimately possible because a massive anomaly-free abelian gauge theory is renormalisable, while the anomalous version is not. This new physics will come in the shape of heavy chiral fermions, to which we turn next.

\section{Anomaly cancellation at the TeV scale} \label{sec:UV}

In this Section we construct an underlying anomaly-free gauge theory that flows to the anomalous $Y_3+a(L_\mu-L_\tau)/6$ model at the weak scale. To do so, 
we extend the SM by a BSM sector containing fermions that are vector-like with respect to the SM gauge symmetry, being chiral only with respect to $U(1)_X$. This is for two important reasons:
\begin{enumerate}
\item We thereby trivially preserve anomaly cancellation within the SM sector itself. 
\item One can Higgs the BSM fermions using scalar fields that are neutral under the SM gauge symmetry. This means their vevs do not break electroweak symmetry prematurely but can safely break $U(1)_X$ at a heavier scale, which also sets the natural mass scale for the BSM fermions.
\end{enumerate}
Note that while our BSM fermions are vector-like with respect to the SM, they cannot be neutral, because their very `purpose' is to cancel the mixed SM-$U(1)_X$ gauge anomalies at the TeV scale where $U(1)_X$ is linearly realised. Specifically, the extra states must be hypercharged in order to cancel the mixed anomaly $\mathcal{A}_{YXX}$. 

Remarkably, one can always restore anomaly cancellation via a pair of chiral fermions, for any value of the parameter $a$. Let $\Psi_1=(\psi_1, \chi_1)$ and $\Psi_2=(\psi_2,\chi_2)$ denote a pair of fermions that are $SU(3)\times SU(2)_L$ singlets, with $\psi_{1,2}$ being their left-handed Weyl components and $\chi_{1,2}$ their right-handed Weyl components. All the Weyl components are taken to have the same hypercharge, denoted $y/6$.\footnote{Since the extra fermions are $SU(2)_L$ singlets, their electric charge will also equal $y/6$.} Under $U(1)_X$, the left-handed components $\psi_{1,2}$ have charges $v_{1,2}/6$, and the right-handed components have charges $w_{1,2}/6$, where
\begin{align} \label{eq:BSMcharges}
v_1 &= 3y+\frac{3a}{2y}-\frac{1}{2},  &w_1 = 3y+\frac{3a}{2y}+\frac{1}{2}, \\
v_2 &= 3y-\frac{3a}{2y}+\frac{1}{2},  &w_2 = 3y-\frac{3a}{2y}-\frac{1}{2}. \nonumber
\end{align}
These charges are integer multiples of one sixth when $a \in (2\mathbb{Z}+1)y$.
For any odd $a$ we can just take $y=1$, while for any even $a$ a suitable value $y=2^k$, $k\in \mathbb{Z}$, can always be found.

These BSM fermions contribute to four anomaly coefficients, namely $\mathcal{A}_{YYX}$, $\mathcal{A}_{YXX}$, $\mathcal{A}_{XXX}$, and $\mathcal{A}_{\mathrm{grav}-X}$. Their contributions to $\mathcal{A}_{YYX}$ and $\mathcal{A}_{\mathrm{grav}-X}$ are proportional to $\sum_i v_i - \sum_i w_i=0$, and so these anomaly coefficients, which recall receive zero contribution from the SM fermions, remain vanishing. On the other hand, the BSM contributions to the two non-linear anomaly coefficients are
\be
(\mathcal{A}_{YXX})^{\mathrm{BSM}}=\frac{y}{6^2}\left(v_1^2 + v_2^2 - w_1^2-w_2^2\right) = -\frac{a}{6} = - (\mathcal{A}_{YXX})^{\mathrm{SM}}, 
\ee
and
\be
(\mathcal{A}_{XXX})^{\mathrm{BSM}}=\frac{1}{6^3}\left(v_1^3 + v_2^3 - w_1^3-w_2^3\right) = -\frac{a}{4} = - (\mathcal{A}_{XXX})^{\mathrm{SM}}, 
\ee
precisely cancelling the anomaly contributions (\ref{eq:gauge anomalies}) from the SM fermions charged under $X=Y_3+a(L_\mu-L_\tau)/6$. We derive the charges (\ref{eq:BSMcharges}) in Appendix~\ref{app:algebra}.

Both $\Psi_1$ and $\Psi_2$ can be given masses via a SM singlet scalar field $\Phi$ with $U(1)_X$ charge $X_\Phi = (w_1-v_1)/6=1/6$. We lift all the extra fermions via the Yukawa interactions
\be
\mathcal{L}^{\Psi}_{\mathrm{mass}} = y_1 \overline{\psi_1} \Phi^\ast \chi_1 +  y_2 \overline{\psi_2} \Phi \chi_2 + \text{h.c.}
\ee 
Once $\Phi$ acquires its non-zero vev $v_\Phi$ as in Eq.~(\ref{eq:vev}), the $U(1)_X$ symmetry is spontaneously broken and the BSM chiral fermions acquires masses
\be \label{eq:fermion_mass}
M_{\Psi_i} = y_i v_\Phi, \; i\in \{1,2\} .
\ee
In \S \ref{sec:pheno} we will find that the natural scale of $U(1)_X$ breaking is at least 5 TeV or so in order to be consistent with the measurement of the electroweak $\rho$-parameter -- specifically, the bound implies $v_X := M_{\ZP}/g_X \geq 4.5$ TeV, while an upper bound from requiring the $\ZP$ couplings are perturbative requires roughly $v_X \leq 15$ TeV.\footnote{In \S \ref{sec:yuk} we will introduce additional scalar spurions charged under $U(1)_X$ that also break the $U(2)^3$ quark flavour symmetries. The vevs of all these spurions will contribute to $M_{\ZP}$, whereas only $v_\Phi$ contributes to the fermion masses (\ref{eq:fermion_mass}).}
We will see in \S \ref{sec:ZPmass} that $v_\Phi$ is expected to be roughly twice $v_X=M_{\ZP}/g_X$, and so the chiral fermion masses are in the range 
\be \label{eq:5TeV}
M_{\Psi_i} \in y_i [10,30] \mathrm{~TeV} \qquad \text{(chiral fermions)}.
\ee
Thus, assuming the Yukawa couplings are no smaller than 0.1, it is natural for the charged fermions $\Psi_{1,2}$ to be anywhere between 1 and 30 TeV, thereby offering a wide target for searches for long-lived charged particles at both the LHC~\cite{Aaboud:2019trc} and future colliders.
We elaborate on this point in the Discussion. In summary, the field content of this model at the TeV scale is recorded in Table \ref{charges}.

\begin{table} 
\begin{centering}
\begin{tabular}{c|c|c|c}
Field & Chirality & $G_{\text{SM}}$ & $6 \times U(1)_X$ \\
\hline
$q_{1,2}$ & L & $(\mathbf{3},\mathbf{2},1/6)$ & $0$ \\
$u_{1,2}$ & R & $(\mathbf{\bar{3}},\mathbf{1},2/3)$ & $0$ \\
$d_{1,2}$ & R & $(\mathbf{\bar{3}},\mathbf{1},-1/3)$ & $0$ \\
\hline
$q_{3}$ & L & $(\mathbf{3},\mathbf{2},1/6)$ & $1$\\
$u_{3}$ & R & $(\mathbf{\bar{3}},\mathbf{1},2/3)$ & $4$ \\
$d_{3}$ & R & $(\mathbf{\bar{3}},\mathbf{1},-1/3)$ & $-2$ \\
\hline
$\ell_1$ & L & $(\mathbf{1},\mathbf{2},-1/2)$ & $0$ \\ 
$\ell_2$ & L & $(\mathbf{1},\mathbf{2},-1/2)$ & $a$ \\
$\ell_3$ & L & $(\mathbf{1},\mathbf{2},-1/2)$ & $-3-a$ \\
$e_1$ & R & $(\mathbf{1},\mathbf{1},-1)$ & $0$ \\
$e_2$ & R & $(\mathbf{1},\mathbf{1},-1)$ & $a$ \\ 
$e_3$ & R & $(\mathbf{1},\mathbf{1},-1)$ & $-6-a$ \\
\hline
$h$ & - & $(\mathbf{1},\mathbf{2},1/2)$ & $3$ \\
\hline
$\psi_1$ & L & $(\mathbf{1},\mathbf{1},y/6)$ & $3y+\frac{3a}{2y}+\frac{1}{2}$ \\
$\psi_2$ & L & $(\mathbf{1},\mathbf{1},y/6)$ & $3y-\frac{3a}{2y}-\frac{1}{2}$ \\
$\chi_1$ & R & $(\mathbf{1},\mathbf{1},y/6)$ & $3y+\frac{3a}{2y}-\frac{1}{2}$ \\
$\chi_2$ & R & $(\mathbf{1},\mathbf{1},y/6)$ & $3y-\frac{3a}{2y}+\frac{1}{2}$ \\
\hline
$\Phi$ & - & $(\mathbf{1},\mathbf{1},0)$ & $1$
\end{tabular}
\caption{\label{charges} The anomaly-free set of $U(1)_X$ charges present at the TeV scale for our model. At energies below the TeV scale the extra Weyl fermions are integrated out, leaving an anomalous $U(1)_X$ gauge symmetry that acts as $X=Y_3+\frac{a}{6}(L_\mu-L_\tau)$ on the SM fermions.  In the final column we record the values of the $U(1)_X$ charges multiplied by six. It is assumed that $a$ is an odd multiple of $y$, which makes all the charges (multiplied by 6) integral in this normalisation. 
}
\end{centering}
\end{table}

\section{Quark masses and mixings} \label{sec:yuk}

In the limit of the renormalisable lagrangian that describes our SM$\times U(1)_X$ model, only the third family fermions have Yukawa couplings by construction,
\be
-\mathcal{L}^{D=4} \supset y_t\,  \overline{q_3} h^c u_3 + y_b\, \overline{q_3} h d_3 + y_\tau\, \overline{\ell_3} h e_3  + \text{h.c.}
\ee
In the quark sector, the renormalisable lagrangian has a global flavour symmetry~\cite{Pomarol:1995xc,Barbieri:1995uv,Barbieri:2011ci,Blankenburg:2012nx,Barbieri:2012uh}
\be
U(2)_q \times U(2)_u \times U(2)_d \times U(1)_{B_3} \equiv U(2)_\mathrm{global}^3 \times U(1)_{B_3},
\ee
where the first two families of each fermion species $(q,\ell,u,d,e)$ transform as doublets under the appropriate factor of $SU(2)_\mathrm{global}^3 \subset U(2)^3_\mathrm{global}$. We sometimes denote these flavour doublets using boldface {\em e.g.} $\bm{q}:=(q_1,q_2)\sim (\bm{2},\bm{1},\bm{1})$. The third family quark fields $q_3$, $u_3$, and $d_3$ are singlets under $U(2)_\mathrm{global}^3$. Here $U(1)_{B_3}$ denotes third family baryon number. (The lepton sector, to which we turn in \S \ref{sec:LFV}, will {\em not} be governed by $U(2)$ flavour symmetries, because these are explicitly broken by the choice of a $U(1)_X$ gauge symmetry that couples to muons but not electrons.)

Going beyond the renormalisable level, there must be further new physics deeper in the UV responsible for generating the light quark masses and mixing angles, such as heavy vector-like fermions.\footnote{This idea dates back to work of Froggatt and Nielsen in the case of $U(1)_X$ breaking spurions~\cite{Froggatt:1978nt,Froggatt:1998he,Leurer:1992wg,Leurer:1993gy}. For the case of $U(2)$-breaking spurions being generated by vector-like fermions, see {\em e.g.}~\cite{Greljo:2018tuh,Bordone:2018nbg} amongst other references.
}
We will not specify an explicit UV sector in this paper, but will parametrize its effects at the TeV scale via an EFT involving the SM fields, the $\ZP$ boson, the anomaly-cancelling chiral fermions $\Psi_{1,2}$, and a set of scalar spurions that we assume encode the breaking of $SU(2)_\mathrm{global}^3 \times U(1)_X$. We remark that $U(2)^3_\mathrm{global}$ quark flavour symmetries have been considered in the context of $\ZP$ models for the NCBAs in {\em e.g.}~\cite{Crivellin:2015lwa,Calibbi:2019lvs,Capdevila:2020rrl}.

In particular, we consider a set of SM singlet scalar spurions in the following representations of $SU(2)_q \times SU(2)_u \times SU(2)_d \times U(1)_X$,
\begin{align} \label{eq:spurions}
&(V_q)^a \sim (\bm{2},\bm{1},\bm{1},-1/6), \qquad &&\Phi \sim (\bm{1},\bm{1},\bm{1},+1/6), \\
&(\Delta_u)^{ab} \sim (\bm{2},\bm{\overline{2}},\bm{1},+1/3), \qquad &&(\Delta_d)^{a\dot{b}} \sim (\bm{2},\bm{1},\bm{\overline{2}},-1/3), \nonumber
\end{align}
where $a$, $b$ and $\dot{b}$ denote fundamental indices for $SU(2)_q$, $SU(2)_u$, and $SU(2)_d$ respectively. This somewhat {\em ad hoc}\footnote{While the particular spurion representations (\ref{eq:spurions}) have been chosen from the `bottom up' to reproduce appropriate Yukawa textures, they might arise naturally in the context of some partial unification of gauge and global symmetries at some high scale. We postpone such investigations for future work. } 
choice mimics the minimal set of spurions first suggested in~\cite{Barbieri:2011ci}, except that here the spurions also transform under $U(1)_X$. The vevs of all the spurions break the $U(1)_X$ gauge symmetry and so contribute to $M_{Z^\prime}$ -- see Section \S \ref{sec:ZPmass}. The global symmetry breaking and the $U(1)_X$ breaking are thus tied together at the same scale. This feature has similarities with the model of~\cite{Falkowski:2015zwa}, in which a particular $U(1)_X$ subgroup of $U(2)$ flavour symmetries was gauged.

For the up-type quarks, one can write down $SU(2)^3_\mathrm{global}\times U(1)_X$-invariant Yukawa terms using the spurions as follows,
\begin{align}
-\mathcal{L} \supset y_t\left[ \overline{q_3} h^c u_3 + x_t(\bm{\overline{q}}\cdot \bm{V_q}) h^c u_3  + w_t\overline{q_3} h^c (\bm{V_q}\cdot \bm{\Delta_u} \cdot \bm{u})\, \Phi^{3} + (\bm{\overline{q}}\cdot \bm{\Delta_u} \cdot \bm{u}) h^c\, \Phi\right],
\end{align}
where $y_t$, $x_t$, and $w_t$ are order-1 coefficients.
Thus, the up-type Yukawa matrix has the form
\be \label{eq:upyuk}
Y_u = y_t
\left(\begin{array}{ccc}
    \Delta_u^{ab} \Phi & x_t V_q^a \\
w_t(\bm{V_q}\cdot \bm{\Delta_u})^b \Phi^3 & 1 \\
  \end{array}\right) \approx
\left(\begin{array}{ccc}
    \Delta_u^{ab} \Phi & x_t V_q^a \\
0 & 1 \\
  \end{array}\right),
\ee
where the approximation holds up to dimension-9 operators that we neglect as small. Similarly for the down-type quarks, we have
\be \label{eq:downyuk}
Y_d = y_b 
\left(\begin{array}{ccc}
    \Delta_d^{a\dot{b}} \Phi & x_b V_q^a \\
w_b(\bm{V_q}\cdot \bm{\Delta_d})^{\dot{b}} \Phi & 1 \\
  \end{array}\right) \approx
y_b
\left(\begin{array}{ccc}
    \Delta_d^{a\dot{b}} \Phi & x_b V_q^a \\
0 & 1 \\
  \end{array}\right),
\ee
again for {\em a priori} order-1 numbers $y_b$, $x_b$, and $w_b$, where this time the approximation holds up to dimension-7 operators that we neglect. We thus expect, assuming that $SU(2)^3_\mathrm{global}\times U(1)_X$ is broken by this set of spurions, a Yukawa structure similar to those studied in {\em e.g.}~\cite{Bordone:2018nbg,Fuentes-Martin:2019mun,Faroughy:2020ina}.

The quark masses and mixings can be well-reproduced by taking the vevs of the spurions, following~\cite{Bordone:2018nbg,Fuentes-Martin:2019mun,Faroughy:2020ina}, to be
\be
\langle \Phi \rangle = \frac{v_\Phi}{\sqrt{2}}, \qquad 
\langle V_q \rangle = \frac{v_V}{\sqrt{2}} \left(\begin{array}{c} 0 \\ 1 \end{array}\right), \qquad
\langle \Delta_f \rangle =\frac{v_f}{\sqrt{2}}  U_f^\dagger
\left(\begin{array}{cc} \delta_f & 0 \\ 0 & 1 \end{array}\right), \qquad f \in \{u,d\},
\ee
where $\Lambda$ is the EFT cutoff scale, $\delta_u \sim m_u/m_c \ll 1$, $\delta_d \sim m_d/m_s \ll 1$, and 
\be
U_f = 
\left(\begin{array}{cc} c_f & s_f e^{i\alpha_f} \\  -s_f e^{-i\alpha_f}  & c_f \end{array}\right), \qquad f \in \{u,d\}.
\ee
If we define small quantities 
\be
\epsilon_A := \frac{v_A}{\sqrt{2}\Lambda}, \qquad A \in \{\Phi, V, u,d\},
\ee
then a good fit to quark masses and mixings is obtained when all the $\epsilon_A$ are $\mathcal{O}(0.1)$, and so we henceforth assume all the $v_A$ are roughly equal.
In \S \ref{sec:ZPmass} we then see that each $v_A \approx 2 v_X$ where $v_X \approx M_{\ZP}/g_X$, for $M_{\ZP}$ the mass of the physical $\ZP$ boson.

As mentioned previously, the combination of electroweak constraints and perturbativity of the $\ZP$ couplings will constrain $v_X$ to lie roughly in the range $[5,15]$ TeV, meaning we expect the vevs $v_A \in [10,30]$ TeV or so. The cutoff scale at which heavy new physics enters should then be 
\be \label{eq:Lambda}
\Lambda \approx (70-210) \mathrm{~TeV}  \qquad \text{(e.g. vector-like fermions)}.
\ee
or so (which should be contrasted with the lighter scale of the chiral fermions (\ref{eq:5TeV}) needed for anomaly cancellation). This is high enough to suppress most contributions of the heavy physics, about which we remain agnostic, to low-energy phenomenology. 

For the most precisely measured low energy processes, notably $\Delta F = 2$ processes such as neutral kaon mixing, the contributions from $\mathcal{O}(100 \mathrm{~TeV})$ scale new physics could still be significant. Suppressing these contributions sufficiently might require further assumptions about the heavy physics. For example, if the heavy states ({\em e.g.} vector-like fermions) are all charged under $U(1)_X$, then operators involving only light quarks are suppressed further by the symmetry breaking spurions. 
In this paper, we are content to postpone such UV dependent considerations for further work, and we will assume in \S \ref{sec:pheno} that the $\ZP$ boson, whose mass is around 3 TeV and whose couplings to the light families are generated by the $U(2)^3$-breaking spurions, dominates the new physics contributions to all low-energy processes including kaon mixing (\S \ref{sec:meson_mixing}).

The Yukawa matrices (\ref{eq:upyuk}) and (\ref{eq:downyuk}) are diagonalized via bi-unitary transformations, {\em viz.} $V_{d_L}^\dagger Y_d V_{d_R} = \text{diag}(y_d,y_s,y_b)$ and $V_{u_L}^\dagger Y_u V_{u_R} = \text{diag}(y_u,y_c,y_t)$, exactly as in~\cite{Bordone:2018nbg,Fuentes-Martin:2019mun}.
One expects small 1-3 mixing angles in both $V_{d_L}$ and $V_{u_L}$, and that the right-handed rotation matrices $V_{d_R}$ and $V_{u_R}$ are especially simple, being approximately equal to just a 2-3 rotation.
For the down-type mixing matrices, which are most important for the phenomenology, an explicit parametrization is~\cite{Bordone:2018nbg,Fuentes-Martin:2019mun}
\be \label{eq:mixing}
V_{d_L}(s_b) = 
\left(\begin{array}{ccc}
	c_d & -s_d  e^{i\alpha_d} & 0 \\ 
	s_d  e^{-i\alpha_d} & c_d & s_b \\ 
	-s_d s_b  e^{-i\alpha_d} & -c_d s_b & 1 \\
  \end{array}\right) ,
\qquad
V_{d_R}(s_b) = 
\left(\begin{array}{ccc}
	1 & 0 & 0 \\ 
	0 & 1 & \frac{m_s}{m_b} s_b \\ 
	0 & -\frac{m_s}{m_b} s_b & 1 \\
  \end{array}\right) 
\approx 
\left(\begin{array}{ccc}
	1 & 0 & 0 \\ 
	0 & 1 & 0 \\ 
	0 & 0 & 1 \\
  \end{array}\right),
\ee
where $s_{d(b)}:=\sin \theta_{d(b)}$ and $c_{d(b)} = \cos \theta_{d(b)}$ for some mixing angles $\theta_d$ and $\theta_b$. The relation $V_{\mathrm{CKM}}=V_{u_L}^\dagger V_{d_L}$ has been used to express the matrix elements of $V_{d_{L/R}}$ in terms of the measured quark masses and mixing angles.
To wit, the 1-2 mixing angle $\theta_d$ is fixed via $s_d/c_d = |V_{td}/V_{ts}|$, and $\alpha_d$ is a phase that is also fixed by the CKM, via $\alpha_d = \text{arg}(V_{td}^\ast V_{ts})$.\footnote{We have set to zero an unconstrained phase in (\ref{eq:mixing}), since it will play no role in our phenomenological analysis.} The final angle $\theta_b$, which corresponds to a mixing angle between the second and third family down-type quarks, is a free parameter in the sense that its value is not fixed by CKM data, but we expect $s_b/c_b =\mathcal{O}(\epsilon_V) =\mathcal{O}(0.1) $. We shall take $\theta_b$ to be a phenomenological parameter of our model in \S \ref{sec:pheno}, which we vary in the vicinity of $|V_{ts}|=0.04$. The NCBA fit as well as the constraint from $B_s-\overline{B_s}$ mixing will depend on $\theta_b$. 

We henceforth drop the terms of order $m_s/m_b$ in (\ref{eq:mixing}), taking the right-handed mixing matrix to be the identity,  $V_{d_R} = \bf{1}$.
We must take $V_{u_L}= V_{d_L} V_{\textrm{CKM}}^\dagger$. Finally, by similar approximations (this time dropping $V_{cb}$-dependent terms that are suppressed by $m_c/m_t$), we can consistently take $V_{u_R} = \bf{1}$ for the purpose of our phenomenological study.

\section{Lepton masses and mixings} \label{sec:LFV}

In the charged lepton sector things are rather different. The $U(2)_\ell \times U(2)_e$ symmetries of the renormalisable Yukawa terms are now explicitly broken by the gauge interactions of the model, because the first and second family leptons (left-handed and right-handed) have different charges under $U(1)_X$. Assuming the same set of spurions as in Eq.~(\ref{eq:spurions}), the structure of the charged lepton Yukawa depends on the (integer-quantized) parameter $a$ that appears in our anomalous gauge symmetry $X=Y_3+a(L_\mu-L_\tau)/6$.
We will see in \S \ref{sec:pheno} that it is phenomenologically reasonable for $a \geq 9$ or so, for a quark mixing angle $\sin \theta_b$ roughly equal to the product of CKM elements $|V_{tb}V_{ts}^\ast|$. 

\subsection{Charged lepton masses}

For general $a\in \mathbb{Z}$, the charged lepton Yukawa matrix is governed by insertions of the $SU(2)^3_\mathrm{global}$-singlet spurion $\Phi$, giving
\be 
Y_e \sim
\left(\begin{array}{ccc}
    \epsilon_\Phi^3 & \epsilon_\Phi^{|a+3|} & \epsilon_\Phi^{|a-3|} \\ 
	\epsilon_\Phi^{|a-3|} & \epsilon_\Phi^3 & \epsilon_\Phi^{|2a+3|} \\ 
	\epsilon_\Phi^{|a+6|} & \epsilon_\Phi^{|2a+6|} & 1 \\
  \end{array}\right).
\ee
For $a \geq 9$, we therefore expect a nearly-diagonal charged lepton Yukawa, 
\be \label{eq:lepton_yuk}
Y_e = 
\left(\begin{array}{ccc}
    c_e \epsilon_\Phi^3 & 0 & 0 \\ 
	0 & c_\mu \epsilon_\Phi^3 & 0 \\ 
	0 & 0 & y_\tau \\
  \end{array}\right) 
+ \mathcal{O}(\epsilon_\Phi^6),
\ee
where $c_e$, $c_\mu$, and $y_\tau$ are dimensionless coefficients that one expects {\em a priori} to be $\mathcal{O}(1)$, with $y_\tau$ being the renormalisable Yukawa coupling of the third family lepton. The model therefore gives an acceptable qualitative prediction $m_{e,\mu} \ll m_{\tau}$, but to fit the observed lepton masses requires $c_e$ and $y_\tau$ to be $\mathcal{O}(0.01)$ (the same is true for $y_b$ on the quark side), which has no explanation within the model.

\subsection{Lepton number symmetries} \label{sec:correlations}

The nearly-diagonal charged lepton Yukawa (\ref{eq:lepton_yuk}) means that the gauge eigenbasis charged lepton states are very closely aligned with the physical mass eigenstates. We therefore predict
\be
V_{e_L} \approx V_{e_R} \approx \bf{1}
\ee
for values of $a$ at least this large, which we assume hereon. This means the $\ZP$ will not mediate LFV at tree-level. We must then take $V_{\nu_L} = V_{\mathrm{PMNS}}^\dagger$ to be consistent with observations of neutrino oscillations. Explaining the origin of this large neutrino mixing is more challenging within the framework we have set up. In Appendix \ref{app:neutrino} we sketch a possible solution.

Even though the $\ZP$ couplings are flavour-conserving it is possible that the heavy physics integrated out at $\Lambda = \mathcal{O}(100 \mathrm{~TeV})$, which is responsible for generating the electron and muon mass via higher-dimensional operators, might give non-neglible lepton flavour violating (LFV) effects. Since the experimental bounds on LFV are so strong, we next estimate the size of these effects using the TeV scale EFT, which is governed by $U(1)_X$ gauge invariance, assuming generic new physics at the scale $\Lambda$.

\subsubsection*{Estimating lepton flavour violation: $\bm{\ell_j \to \ell_i \gamma}$}

First we estimate the size of the ($\mu\to e\gamma$)-inducing dipole operators
\be
\mathcal{L} \supset \frac{1}{\Lambda_{e\mu}^2}\frac{v}{\sqrt{2}} \overline{e} \sigma^{\mu\nu} P_X \mu F_{\mu\nu}+\mathrm{h.c.}
\ee
within our model, where $e$ and $\mu$ denote the electron and muon fields below the scale of electroweak symmetry breaking, and $P_X$ is some chirality operator. The experimental limit on $BR(\mu\to e\gamma)$ translates to a limit on the heavy scale of $\Lambda_{e\mu} \gtrsim 58\, 000$ TeV, as constrained by the full MEG dataset~\cite{TheMEG:2016wtm}.

This lagrangian descends from the following operators in the TeV scale EFT,
\begin{align}
\mathcal{L} \supset &\frac{1}{\Lambda^2} \left(\overline{\ell_1} \sigma^{\mu\nu} e_2\right) \left(c_{1} \tau^a h W^{a}_{\mu\nu}+c_{2}h F^Y_{\mu\nu} \right) \left(\frac{\Phi^\dagger}{\Lambda}\right)^{a+3} \nonumber \\
+ &\frac{1}{\Lambda^2} \left(\overline{\ell_2} \sigma^{\mu\nu} e_1\right) \left(c_{3} \tau^a h W^{a}_{\mu\nu}+c_{4} h F^Y_{\mu\nu} \right) \left(\frac{\Phi}{\Lambda}\right)^{a-3} + \mathrm{h.c.},
\end{align}
where $W_{\mu\nu}^a$ and $F^Y_{\mu\nu}$ denote the field strengths for $SU(2)_L$ and $U(1)_Y$ (of which the photon field strength is a linear combination), and the $c_i$, $i=1,\dots,4$ denote unknown Wilson coefficients.
Taking the larger contribution from the second line, the $\mu \to e\gamma$ decay therefore roughly requires 
\be
\frac{\Lambda}{\sqrt{\tilde{c}}} \epsilon_\Phi^{-\frac{a-3}{2}} \gtrsim 58\, 000 \mathrm{~TeV},
\ee
where $\tilde{c}$ denotes a characteristic scale of the contributing Wilson coefficients $c_3$ and $c_4$.
For, say, $\Lambda \approx 100$ TeV, and recalling $\epsilon_\Phi \approx 0.1$, this is satisfied for the phenomenologically reasonable range $a \geq 9$ without any need for the Wilson coefficients to be small.

A similar suppression is predicted for $\tau \to e\gamma$, again for generic $\mathcal{O}(1)$ coefficients in the TeV scale EFT -- but the experimental limit on the branching ratio is a factor $10^5$ looser than for $\mu \to e\gamma$. The prediction for LFV in $\tau \to \mu \gamma$ is truly tiny, with the associated scale being $\Lambda_{\tau \mu} \sim \Lambda \epsilon^{-(2a+3)/2} \gtrsim 10^{12}$ TeV. So we see that gauging a high quality lepton number symmetry, as follows the choice $X=Y_3+a(L_\mu-L_\tau)/6$, means the contributions of the heavy $\Lambda$-scale physics to $\ell_j \to \gamma \ell_i$ decays is pushed back to such heavy scales as to be phenomenologically irrelevant.

\subsubsection*{Estimating lepton flavour violation: $\bm{\ell_j \to 3\ell_i}$}

We now estimate the natural size of flavour-violating four-lepton operators in the TeV scale EFT, assuming generic $\Lambda$-scale physics with $\mathcal{O}(1)$ Wilson coefficients. These operators will disturb the branching ratios for $\mu \to 3 e$, $\tau \to 3 e$, and $\tau \to 3 \mu$. For $\mu \to 3 e$, the relevant four-lepton operators come from
\be
\mathcal{L} \supset \frac{c_{LL}}{\Lambda^2} \left(\overline{\ell_1} \gamma^\mu \ell_2 \right)\left(\overline{\ell_2} \gamma^\mu \ell_2 \right) \left(\frac{\Phi^\dagger}{\Lambda}\right)^{a}
+ \frac{c_{LR}}{\Lambda^2} \left(\overline{\ell_1} \gamma^\mu \ell_2 \right)\left(\overline{e_2} \gamma^\mu e_2 \right) \left(\frac{\Phi^\dagger}{\Lambda}\right)^{a} + \dots
\ee
The new physics contribution to $BR(\mu \to 3 e)$ from this interaction is of order
\be
\Delta BR(\mu \to 3 e) \sim \frac{m_\mu^5}{768\pi^3 \Gamma_\mu} \frac{1}{\Lambda^4} \epsilon^{2a} \lesssim 10^{-29}
\ee
or so, for $\Lambda \approx 100$ TeV, $\epsilon \approx 0.1$, and $a \geq 9$ as usual, and for $\mathcal{O}(1)$ Wilson coefficients. This contribution is far smaller than the experimental bound on the branching ratio, which is of order $10^{-12}$. The corresponding contributions to LFV decays of tauons to three charged leptons are expected to be far smaller still, thanks to even more $\Phi$ insertions in the EFT.

The upshot of these estimates is that generic heavy physics at $\mathcal{O}(100 \mathrm{~TeV})$, which recall is needed to explain the light Yukawas, is expected to give very small contributions to LFV processes because of the structure imposed by $U(1)_X$ invariance. Given also that the $\ZP$ couplings to charged leptons are very nearly diagonal, based on Eq.~(\ref{eq:lepton_yuk}), the model possesses high quality lepton number symmetries. Recall that preserving these accidental symmetries of the SM, which are borne out so strongly in Nature, was one of the driving motivations behind the model introduced in this paper.

This kind of mechanism, in which the lepton number global symmetries are underwritten by a family-dependent gauge symmetry, has been previously suggested as a natural way of explaining the observed LFUV in the NCBAs without any expectation of accompanying LFV (see {\em e.g.}~\cite{Altmannshofer:2019xda} for other $Z^\prime$ examples, and {\em e.g.}~\cite{Davighi:2020qqa,Greljo:2021xmg} for examples involving scalar leptoquarks). 

Finally, we emphasize that all of the LFV effects in our model become smaller still as we dial the parameter $a$ (and thus the $U(1)_X$ charges of the anomaly-cancelling BSM fermions) to larger values. As we show in \S \ref{sec:pheno}, for reasonable quark mixing angles and $\ZP$ masses, $a$ can be as large as 50 or so before the $\ZP$ becomes non-perturbative. 
Conceptually, by allowing the $U(1)_X$ gauge anomalies to be cancelled via heavy chiral fermions, we can effectively decouple the magnitude of the lepton charges from the magnitude of the quark charges through the parameter $a$ (where $a$ can be thought of as the degree of mismatch between the quark and lepton contributions to anomaly cancellation). This feature, which is in contrast to anomaly-free $U(1)_X$ models in which the ratios of SM charges are typically fixed, allows one to push the lepton charges much higher than the quark charges, which in turn allows us to (i) preserve lepton number symmetries to a very high quality, and (ii) explain the $B$ anomalies while suppressing other effects {\em e.g.} in $B_s-\overline{B}_s$ mixing and in the electroweak $\rho$-parameter -- as we discuss in detail in the following Section.

\section{Phenomenological analysis} \label{sec:pheno}

Having described the theoretical motivations for an anomalous $Y_3 + a(L_\mu-L_\tau)/6$ model in previous Sections, together with various model building details, we now turn to the low-energy phenomenology associated with the $\ZP$ boson. To do so, we begin by first writing down the BSM couplings of the (low-energy limit of the) model more explicitly. The model shares many features, such as $Z-\ZP$ mixing, with third family hypercharge models~\cite{Allanach:2018lvl,Allanach:2019iiy}. We then compute what we expect to be the dominant constraints on the model, saving a more comprehensive phenomenological analysis for future work. We neglect all the (uncalculable) contributions from the unspecified heavy physics at $\mathcal{O}(100\mathrm{~TeV})$ in what follows, assuming the $\ZP$ contributions to dominate the low-energy phenomenology.

\subsection{Details of the model}

The field content of the model, including the charges under $U(1)_X$, are specified in Table \ref{charges}. We take the mixing matrices $V_{d_L}$, $V_{u_L}$, {\em etc.}, which specify the misalignment between the interaction basis and the mass basis for fermions, to be those given in \S\S \ref{sec:yuk} and \ref{sec:LFV}. Recall that the forms of these mixing matrices were consistent with an EFT analysis of effective Yukawa couplings that could be written down using insertions of a minimal set of spurions.

\subsubsection{Symmetry breaking: the $\ZP$ mass and mixing} \label{sec:ZPmass}

The vevs of all four $SU(2)^3_\mathrm{global}\times U(1)_X$ breaking spurion fields contribute to the mass of the $\ZP$ boson. Assuming that the vevs of these spurions are all much greater than the Higgs vev $v$, the tree-level mass formula is
\be
M_{\ZP} \approx \frac{g_X}{6}\sqrt{v_\Phi^2 + v_V^2 +4v_u^2+4v_d^2} =: g_X v_X,
\ee
where the second equality defines $v_X$ as a phenomenological parameter indicative of the scale of $U(1)_X$ breaking. 
Note that $v_\Phi$ is independently an important parameter phenomenologically, because it sets the mass scale (\ref{eq:5TeV}) of the anomaly cancelling charged fermions. We saw in \S \ref{sec:yuk} that the quark Yukawas are well-modelled if all the spurion vevs are roughly equal (with $\epsilon_A := v_A/\sqrt{2}\Lambda\approx 0.1$), so we expect $v_\Phi \approx 6v_X/\sqrt{10}\approx 2 v_X$, as was used in (\ref{eq:5TeV}).

Most of the bounds that we will compute constrain the combination $v_X = M_{\ZP}/g_X$, since the effects of integrating out the $\ZP$ are all suppressed by powers of this ratio. The only constraints on $M_{\ZP}$ directly are that it ought to be greater than 2 TeV or so to evade the LHC direct searches, and in \S \ref{sec:pheno} we will find that it cannot be much heavier than 6 TeV without its couplings becoming non-perturbative.

The fact that the Higgs field $h$ is charged under $U(1)_X$ leads to $Z-\ZP$ mixing, which is exactly the same as in~\cite{Allanach:2018lvl,Allanach:2019iiy}.  
Specifically, the electrically neutral gauge bosons $B_\mu$, $W^3_\mu$ and $X_\mu$ are related to the physical mass eigenstates $\gamma$, $Z^0$ and $Z^\prime$ via the linear transformation
\be
\left(\begin{array}{c}
	B_\mu \\ 
	W^3_\mu \\ 
	X_\mu \\
  \end{array}\right) 
= 
\left(\begin{array}{ccc}
	c_w & -s_w c_z & s_w s_z \\ 
	s_w & c_w c_z & -c_w s_z \\ 
	0 & s_z & c_z \\
  \end{array}\right)
\left(\begin{array}{c}
	A_\mu \\ 
	Z_\mu \\ 
	\ZP_\mu \\
  \end{array}\right) ,
\ee
where $c_{w(z)}:=\cos \theta_{w(z)}$ and $s_{w(z)}:=\sin \theta_{w(z)}$, where $\theta_w$ is the Weinberg angle (such that $\tan \theta_w = g^\prime/g$ is the ratio of $U(1)_Y$ to $SU(2)_L$ gauge couplings), and the $Z-\ZP$ mixing angle $\theta_z$ is given by
\be
\sin \theta_z \approx \frac{g_X}{\sqrt{g^2+g^{\prime 2}}} \left(\frac{M_Z}{M_{\ZP}} \right)^2\, .
\ee
This mixing means, most importantly, that the SM $Z$ boson picks up family non-universal couplings to leptons from mixing with the $X$ gauge field, which are constrained by LEP measurements (see below). The invisible width of the $Z$ boson is also altered.

\subsubsection{$\ZP$ couplings to fermions}

Letting $\bm{d}_L := (d_L, s_L, b_L)^T$ {\em etc.} now denote the 3-component column vectors of mass basis fermion fields, the $\ZP$ couplings to fermions are given by
    \begin{eqnarray}
      {\mathcal L}_{\psi X} =
      &-g_{X} &\left( 
      \frac{1}{6}{\overline{{\bm u_L}}} \Lambda^{(u_L)} \slashed{X} {\bm
        u_L} + \frac{1}{6}{\overline{{\bm d_L}}} \Lambda^{(d_L)} \slashed{X} {\bm
        d_L} +\frac{2}{3}\overline{t_R} \slashed{X} t_R
           -\frac{1}{3} \overline{b_R} \slashed{X} b_R
           \right.  \label{eq:Xcouplings} \\ && \left.
         +\frac{a}{6} \overline{\mu} \slashed{X} \mu - \frac{3+a}{6}\overline{\tau_L} \slashed{X} \tau_L - \frac{6+a}{6}\overline{\tau_R} \slashed{X} \tau_R +
\frac{a}{6}{\overline{{\bm \nu_L}}} \Lambda^{(\nu_L)} \slashed{X} {\bm \nu_L} 
                    \right) + \mathcal{O}\left(M_Z/M_{\ZP}\right)^2,  \nonumber
    \end{eqnarray}
where the matrices $\Lambda^{(d_L)}$, $\Lambda^{(u_L)}$, and $\Lambda^{(\nu_L)}$ are given by
\begin{align}
\Lambda^{(d_L)} &:= V_{d_L}^\dagger \xi V_{d_L}, \quad &&\xi = \textrm{diag}(0,0,1), \\
\Lambda^{(u_L)} &:= V_{u_L}^\dagger \xi V_{u_L}, \\
\Lambda^{(\nu_L)} &:= V_{\textrm{PMNS}} \Omega V_{\textrm{PMNS}}^\dagger, \quad &&\Omega = \textrm{diag}\left(0,1,-(3+a)/a\right). \label{eq:nu_coup}
\end{align}
Recall $V_{d_L}$, which is a function of a single undetermined mixing angle $\theta_b$, is defined in (\ref{eq:mixing}), and that 
$V_{u_L}= V_{d_L} V_{\textrm{CKM}}^\dagger$. The most important couplings, phenomenologically, are those of the $\ZP$ to the down-type quarks, so it is useful to write them out explicitly:
\be \label{eq:dtype_coup}
\Lambda^{(d_L)} =
\left(\begin{array}{ccc}
	s_d^2 s_b^2 & s_d s_b^2 e^{i\alpha_d} & -s_d s_b e^{i\alpha_d} \\ 
	s_d s_b^2 e^{-i\alpha_d} & c_d^2 s_b^2 & -c_d s_b \\ 
	-s_d s_b e^{-i\alpha_d} & -c_d s_b & 1 \\
  \end{array}\right).
\ee
As expected based on third family alignment of the $X$ boson couplings in the interaction basis, the couplings to the light family quarks are suppressed by powers of the small mixing angles $\theta_d$ and $\theta_b$, where recall that $\tan \theta_d = |V_{td}/V_{ts}| = 0.210 \pm 0.001 \pm 0.008$~\cite{Zyla:2020zbs}, whereas $\theta_b$ is a parameter of the model. Note that the phases are such that $\mathrm{arg}(\Lambda^{(d_L)})_{12}= \mathrm{arg}(V_{td}^\ast V_{ts})$ and $\mathrm{arg}(\Lambda^{(d_L)})_{13}\approx \mathrm{arg}(V_{td}^\ast V_{tb})$.

\subsection{$B$-decays and related constraints}

We now compute the leading constraints on the model, as functions of its four parameters 
\be
\mathrm{Parameters:~} \, \{g_X, M_{\ZP}, a,  \theta_b\},
\ee
and identify regions of parameter space that give a good fit to the NCBA data while agreeing with the other constraints (see Figures~\ref{fig:const} and~\ref{fig:pert}).

\subsubsection{The $b\to s \ell\ell$ anomalies}
Using (\ref{eq:Xcouplings}) and (\ref{eq:dtype_coup}), the couplings of the $\ZP$ relevant to the NCBAs are
\be \label{eq:Lag_Xcouplings}
\mathcal{L}_{\psi X} \supset -g_\mu \overline{\mu} X^\rho \gamma_\rho \mu - g_{sb}(\overline{s} Z^\prime_\rho \gamma^\rho P_L b + \mathrm{h.c.}), \qquad  g_{sb} = -c_d s_b g_X/6, \quad g_{\mu} = g_X a/6.
\ee
Upon integrating out the $\ZP$ boson and matching the SMEFT coefficients to the WET coefficients at the scale of $B$-mesons, these couplings give rise to the low-energy lagrangian 
\be
\mathcal{L}_\text{WET} \supset \mathcal{N}\Delta C_9^{bs\ell\ell}\mathcal{O}_9^{bs\ell\ell}, \qquad
\mathcal{O}_9^{bs\ell\ell} = (\overline{s} \gamma_\mu P_L b)(\overline{\ell}\gamma^\mu \ell),
\ee
where $\Delta$ denotes the new physics contributions only, and the normalisation factor is $\mathcal{N} =  \frac{4 G_F}{\sqrt{2}} V_{tb} V_{ts}^\ast \frac{e^2}{16 \pi^2}$.
In terms of our model parameters, the contribution to $\Delta C_9^{bs\mu\mu}$ is
\be \label{eq:C9 formula}
\Delta C_9^{bs\mu\mu} 
= a g_X^2 \frac{c_d s_b}{V_{tb} V_{ts}^\ast} \frac{1.45 \text{~TeV}^2}{M_{Z^\prime}^2}.
\ee
A recent  pure $\Delta C_9^\mu$ global fit to the rare $B$-decay data~\cite{Altmannshofer:2021qrr}, that takes into account the updated LHCb measurements (from March 2021) of $R_K$~\cite{Aaij:2021vac} and $BR(B_s\to \mu^+\mu^-)$~\cite{LHCb_Bsmumu}, finds $\Delta C_9^\mu = -0.82^{+0.14}_{-0.14}$. Noting that (the real part of) $V_{ts}$ is negative, this requires $s_b a > 0$, and translates to the following 2$\sigma$ range on our model parameters, 
\be
1.15 \text{~TeV} \leq \frac{M_{Z^\prime}}{g_X  \sqrt{a}}\sqrt{\frac{|V_{tb} V_{ts}^\ast|}{c_d s_b}} \leq 1.64 \text{~TeV}.
\ee

\subsubsection{$\Delta F = 2$ neutral meson mixing} \label{sec:meson_mixing}

As for any $Z^\prime$ model for the NCBAs, there is a tree-level contribution to $B_s-\overline{B}_s$ meson mixing, which favours a smaller coupling of the $\ZP$ to the quark current $b\bar{s}$. We expect this, along with related $\Delta F = 2$ amplitudes, to provide an important constraint on the model.
The 2$\sigma$ bound from $B_s-\bar{B}_s$ mixing is~\cite{DiLuzio:2019jyq}
\be
0.89 \leq \left| 1+ \frac{C_{bs}^{LL}}{R_\text{SM}} \right| \leq 1.09,
\ee
where in our model
\be \label{eq:Bs_coeff}
C_{bs}^{LL} = \frac{1}{4\sqrt{2} G_F M_{Z^\prime}^2} \left( \frac{g_{sb}}{V_{tb} V_{ts}^\ast }\right)^2 = \frac{g_X^2}{144 \sqrt{2} G_F M_{Z^\prime}^2} 
\left( \frac{c_d s_b}{V_{tb} V_{ts}^\ast} \right)^2.
\ee
The SM loop factor here is $R_\text{SM} = (1.310 \pm 0.010) \times 10^{-3}$. Comparing with the upper limit therefore gives the bound
\be
\frac{M_{Z^\prime}}{g_X}
\left| \frac{V_{tb} V_{ts}^\ast}{c_d s_b} \right| 
\geq 1.9 \text{~TeV}
\ee
at 2$\sigma$, neglecting the small imaginary part of $V_{ts}$.

There are also $s_b$-dependent bounds coming from lighter neutral meson mixing measurements, specifically in the kaon and $B_d$ systems. For both systems we assume the tree-level $\ZP$ contribution, which we can calculate explicitly, dominates over the loop-induced contributions from the heavy physics at the scale $\Lambda = \mathcal{O}(100\mathrm{~TeV})$, which we cannot calculate without an explicit model of the new physics.

\begin{figure}
\begin{center}
\unitlength=15cm
\begin{picture}(1,0.45)(0,0)
    \put(-0.25,-0.05)
{\includegraphics[width=0.85\textwidth]{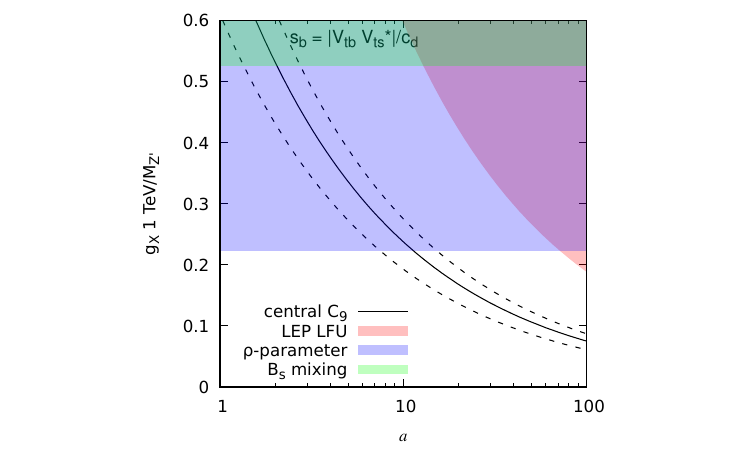}}
    \put(0.3,-0.05)
{\includegraphics[width=0.85\textwidth]{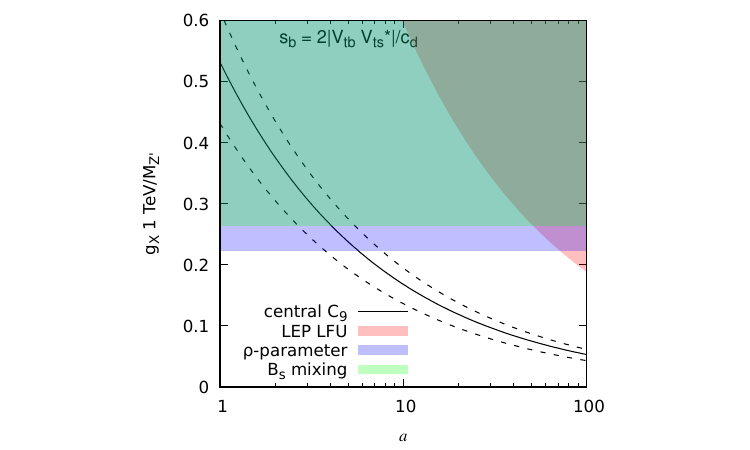}}
\end{picture}
\vspace{0.3in}
\caption{\label{fig:const} Summary of constraints on the $X=Y_3+\frac{a}{6}(L_\mu-L_\tau)$ model introduced in this paper, in the $g_X/M_{\ZP}$ {\em vs.} $a$ plane, for two values of the 23 down-type quark mixing angle $\theta_b$. In both plots, the white region between the dashed line contours is allowed at $95\%$ CL\@. In addition to the band that fits the NCBAs, we show the constraints ($95\%$ CL\@) from $B_s-\bar{B}_s$ mixing, LEP LFU, and the $\rho$-parameter. For the larger value of the mixing angle (right-hand plot), the $B_s-\bar{B}_s$ mixing and the $\rho$-parameter constraints are of a similar strength, allowing one to access the widest range of values for the parameter $a$. In Fig.~\ref{fig:pert} we also show the constraint on the gauge coupling from requiring perturbative unitarity. }  
\end{center}
\end{figure}

From $B_d-\overline{B}_d$ mixing, we use~\cite{Charles:2013aka} to calculate the bound
\be
\frac{M_{Z^\prime}}{g_X}
\left| \frac{V_{tb} V_{td}^\ast}{s_d s_b} \right| 
\geq 1.2 \text{~TeV}.
\ee
Similarly, from kaon mixing~\cite{Charles:2013aka}, the bound is
\be
\frac{M_{Z^\prime}}{g_X}
\left| \frac{V_{ts} V_{td}^\ast}{s_d s_b^2} \right| 
\geq 0.68 \text{~TeV}.
\ee
To compare the strengths of these three bounds, it is convenient to parametrize the mixing angle $s_b$ via the ratio
\be \label{eq:kappa}
\kappa := \frac{c_d s_b}{|V_{tb} V_{ts}^\ast|}.
\ee
Then the three bounds can be written compactly as
\be
\frac{M_{Z^\prime}}{\kappa g_X} \geq 
\begin{cases}
1.9 \text{~TeV}, \qquad &(B_s) \\
1.2 \text{~TeV}, \qquad &(B_d) , \\
\kappa\; 0.68 \text{~TeV}, \qquad &(\mathrm{Kaon}).
\end{cases}
\ee
Thus, the $B_d$ mixing constraint is everywhere weaker than the $B_s$ mixing constraint and so can be neglected. The kaon mixing constraint is generally the weakest, but its relative importance grows with $\kappa$, {\em i.e.} grows with the mixing angle $\theta_b$. Indeed, kaon mixing becomes more constraining than $B_s$ mixing for $\kappa > 2.8$. That said, we will be most interested in regions of parameter space with $\kappa < 2.3$ (see footnote~\ref{foot:kappa}), for which all the neutral meson mixing constraints are weaker than a constraint from the electroweak $\rho$-parameter (below). In Fig.~\ref{fig:const}, we fix $\kappa=1$ (left plot) and $\kappa=2$ (right plot), while for Fig.~\ref{fig:pert} we have $\kappa=2$ for both plots. 

\subsubsection{Some electroweak constraints}

\paragraph{LFU of \boldmath{$Z$} boson couplings:}
The $Z-\ZP$ mixing described above, which inevitably follows from charging the Higgs under $U(1)_X$, gives a small BSM contribution to the $Z$ boson coupling to muons but not electrons, thus violating LFU in $Z$ boson decays. This is tightly constrained by the observable $R\equiv\frac{\Gamma(Z\rightarrow e^+e^-)}{\Gamma(Z\rightarrow \mu^+\mu^-)}$, which LEP measured to be $R_{\text{LEP}} =0.999\pm 0.003$~\cite{Zyla:2020zbs}.
In our model this LFUV observable is computed as the ratio of partial widths,
\begin{equation}
R_{\text{model}} = \frac{|g_Z^{e_L e_L}|^2+ |g_Z^{e_R e_R}|^2}{|g_Z^{\mu_L \mu_L}|^2+|g_Z^{\mu_R \mu_R}|^2} 
\end{equation}
where $g_Z^{ff}$ is the coupling of the physical $Z$ boson to the fermion
anti-fermion pair
$f\bar f$. While the $Z$ couplings to electrons as unchanged from their SM values, the muon couplings are altered due to the $Z-\ZP$ mixing.
To leading order in $\sin\theta_z$, we have 
\begin{eqnarray}
\begin{aligned}
g_Z^{\mu_L \mu_L} &=-\frac{1}{2}g\cos\theta_w+\frac{1}{2}g'\sin\theta_w+\frac{a}{6}g_X\sin\theta_z,\\
g_Z^{\mu_R \mu_R} &= g'\sin\theta_w + \frac{a}{6}g_X\sin\theta_z.
\end{aligned}
\end{eqnarray}
Expanding $R_{\text{model}}$ to leading order in $\sin\theta_z$, we obtain
\begin{eqnarray}
R_{\text{model}} = 1
+\frac{2a}{3}g_X\frac{(g\cos\theta_w-3g'\sin\theta_w)\sin\theta_z}{(g\cos\theta_w-g'\sin\theta_w)^2+4{g'}^2\sin^2\theta_w}
= 1+0.187 a g_X^2 \left(\frac{M_Z}{M_{Z^\prime}}\right)^2. 
\end{eqnarray}
Comparing with the LEP measurement gives the  2$\sigma$ constraint
\be
\frac{M_{Z^\prime}}{g_X  \sqrt{a}} \geq 0.53 \text{~TeV}.
\ee
Together with the bound from fitting the NCBAs, this puts a lower limit on the size of the quark mixing angle $\theta_b$, which is
\be
\kappa \geq 0.21 \, ,
\ee
where $\kappa$ was defined in (\ref{eq:kappa}).
As long as we choose a mixing angle that is at least this big, the LEP LFU constraint is satisfied everywhere in the 2$\sigma$ range that fits the NCBAs.

\paragraph{The \boldmath{$\rho$}-parameter:}
Charging the Higgs under $U(1)_X$ also leads to a constraint from the precisely measured $\rho$-parameter, along with other electroweak observables. A proper analysis of these electroweak precision observables requires a global fit, as recently performed for a related family of models in~\cite{Allanach:2021kzj}.\footnote{Indeed, in these studies it was found that third family hypercharge models fit the $\rho$-parameter somewhat {\em better} than the SM does, easing a $\sim 2\sigma$ discrepancy between the $W$ boson mass and its SM prediction.} Such a global fit is beyond the scope of the preliminary phemonological study of the $Y_3+a(L_\mu-L_\tau)/6$ model that we present in this paper. Instead, we estimate the bound from the $\rho$-parameter in isolation, using simplified fits to electroweak data.
To leading order in $M_Z^2/M_{Z'}^2$ the model predicts, as in~\cite{Davighi:2020nhv},
\begin{equation} \label{eq:rho_model}
\rho_0 = 1+\frac{ g_X^2}{g^2+g'^2} \frac{M_Z^2}{M_{Z^\prime}^2}.
\end{equation}
A global fit to determine $\rho_0$, $M_H$, $M_t$, and $\alpha_s(M_Z)$ gives
$\rho_0 = 1.00039 \pm 0.00019$~\cite{Zyla:2020zbs}.
Because the BSM contribution (\ref{eq:rho_model}) is independent of both the $b\bar{s}$ and the muon coupling, depending only on the $\ZP$ coupling to the Higgs, this results on a straightforward bound on the mass over the coupling of the $\ZP$,
\be
\frac{M_{Z^\prime}}{g_X} \geq 4.5 \text{~TeV},
\ee
that is independent of the parameters $\theta_b$ and $a$. 

\paragraph{Invisible width of the \boldmath{$Z$} boson:}
Because of the $Z-\ZP$ mixing, the couplings of the $Z$ boson to left-handed neutrinos are altered, resulting in a shift in the invisible width of the $Z$ boson. The couplings of the $X$ boson to neutrinos are given in the mass eigenbasis in Eqs. (\ref{eq:Xcouplings}, \ref{eq:nu_coup}). But, since the collider measurements of $Z$ decays to invisibles are not sensitive to neutrino flavour, we can compute the shift in the invisible width using the interaction basis couplings of the $X$ (which are not cluttered by the PMNS-induced rotation matrices). Denoting the neutrino interaction eigenstates with primed labels, the relevant couplings are
\begin{equation}
    {\mathcal L}_{\bar \nu \nu Z} =-
     \frac{g}{2 c_w} \overline{{\nu_L'}_e} \slashed{Z} P_L {\nu_L'}_e
    - \overline{{\nu_L'}_\mu}\left( \frac{g}{2 c_w} + \frac{a}{6} g_X
    s_z \right)\slashed{Z} {\nu_L'}_\mu
    - \overline{{\nu_L'}_\tau}\left( \frac{g}{2 c_w} - \frac{a+3}{6} g_X
    s_z \right)\slashed{Z} {\nu_L'}_\tau. \label{Znunu}
  \end{equation}
Summing over the partial widths $\Gamma_{\nu_i} = |g_{\nu'_i}|^2 M_Z/(24\pi)$ for each decay, we find the correction
\be
\frac{\Delta \Gamma_{\text{inv}}^{(\ZP)}}{M_Z} = - \frac{g_X^2}{48\pi} \left( \frac{M_Z}{M_{\ZP}} \right)^2
\ee
(which happens to coincide with the correction found in~\cite{Allanach:2019iiy} for the `deformed Third Family Hypercharge model'.)

The experimental measurement of the invisible $Z$ width is in fact slightly below the SM prediction, with $\Delta \Gamma_{\text{inv}}^{(\text{exp})} - \Delta \Gamma_{\text{inv}}^{(\text{SM})} = -2.5 \pm 1.5$ MeV. So, provided the mass ratio $M_Z/M_{\ZP}$ is small enough, the $Z-\ZP$ mixing in our model can in fact ease this slight tension between the SM and the data. As a constraint, this measurement implies 
\be
\frac{M_{Z^\prime}}{g_X} \geq 0.95 \text{~TeV},
\ee
a very weak constraint on the model which does not feature in any of our plots.

\subsubsection{Combined constraints} 

We plot the combination of the first four constraints discussed in this Section, which we expect to be the dominant ones for our model, in Fig.~\ref{fig:const}. We do so for two different values of the mixing angle $\theta_b$. All of the constraints that we have computed depend on $g_X$ and $M_{\ZP}$ only via their ratio, and so by fixing $\theta_b$ we can visualise the constraints via 2d plots with the parameter $a$ on the abscissa.

To overcome the constraint from the $\rho$-parameter, while fitting the NCBAs, favours larger values of the product $a\sin\theta_b$ -- but the mixing angle $\theta_b$ cannot be increased too high without contravening the bounds from neutral meson mixing. For example, we see from the left plot of Fig.~\ref{fig:const} that, for $\sin\theta_b = |V_{tb} V_{ts}^\ast|/c_d$, the $\rho$-parameter constraint requires $a \geq 12$ (remembering that, in our normalisation, $a$ is an integer). But if we increase the quark mixing angle by a factor of two (right plot of Fig.~\ref{fig:const}), the $B_s-\bar{B}_s$ mixing and the $\rho$-parameter constraints become comparable and both can be satisfied, while fitting the NCBAs, for $a \geq 6$.\footnote{If we increase the mixing angle much further, the $B_s-\bar{B}_s$ mixing constraint becomes more important than the electroweak constraint, and, because the $B_s-\bar{B}_s$ mixing depends directly on the mixing angle, no smaller values of $a$ can be accessed. \label{foot:kappa}}

\subsection{$Z^\prime$ decays and width} \label{sec:decay}

Of course, if a $\ZP$ boson like the one detailed in this paper were indeed the explanation behind the observed discrepancies in $b\to s \ell\ell$ transitions, then one would hope to produce the $\ZP$ boson in $pp$ collisions at the LHC. The ATLAS and CMS experiments have published a variety of searches for such a heavy gauge boson, notably in decay channels $\ZP \to t\bar{t}$~\cite{Aaboud:2018mjh,Aaboud:2019roo} (36 fb$^{-1}$ of 13 TeV data), $\ZP \to \tau^+\tau^-$~\cite{Aad:2015osa} (10 fb$^{-1}$ of 8 TeV data), and, most promisingly, $\ZP \to \mu^+\mu^-$~\cite{Aad:2019fac} (139 fb$^{-1}$ of 13 TeV data). 

\begin{figure}
\begin{center}
\unitlength=15cm
\begin{picture}(1,0.45)(0,0)
    \put(-0.25,-0.05)
{\includegraphics[width=0.85\textwidth]{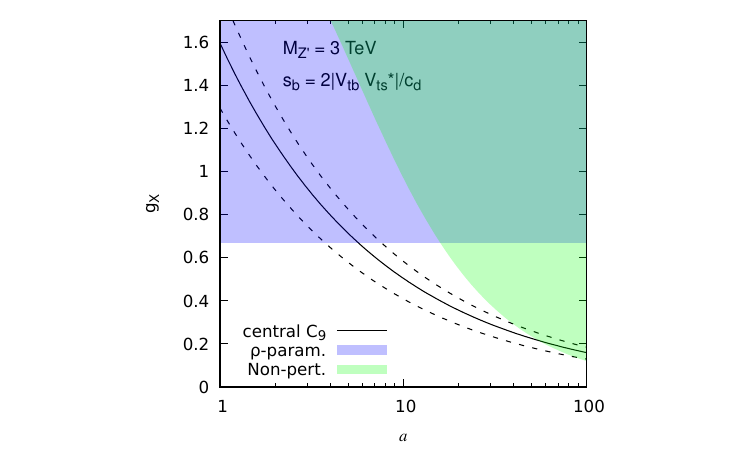}}
    \put(0.3,-0.05)
{\includegraphics[width=0.85\textwidth]{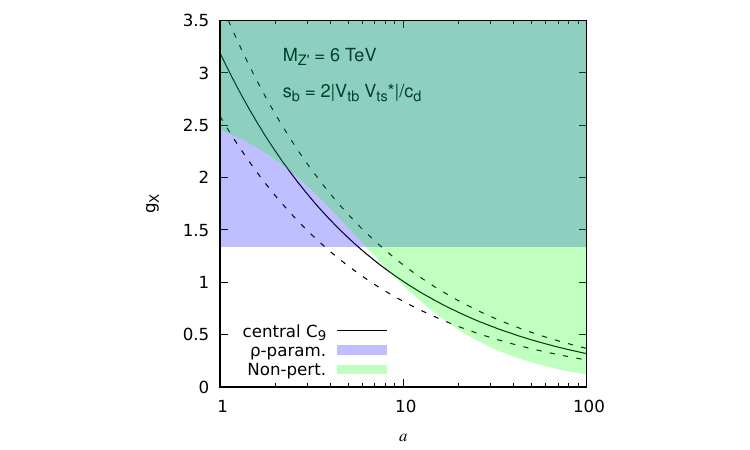}}
\end{picture}
\vspace{0.3in}
\caption{\label{fig:pert} Constraints on the gauge coupling $g_X$, as a function of the parameter $a$, from demanding perturbative unitarity (specifically, from requiring the $\ZP$ width is no more than one third its mass). We plot the perturbative unitarity bound, together with the constraints from fitting the NCBA data and from the electroweak $\rho$-parameter, for two different values of the $\ZP$ mass; 3 TeV (left) and 6 TeV (right). 
In both plots, the white region between the dashed line contours is allowed at $95\%$ CL\@. While there is a wide region of available parameter space fitting the NCBAs for a 3 TeV $\ZP$ boson, with the parameter $a$ ranging between 6 and 50, for the heavier 6 TeV $\ZP$ it is only just possible to fit the NCBA data, and so the $\ZP$ mass cannot be pushed much higher (for any value of $a$) without it becoming non-perturbative. In both plots, the constraint from $B_s-\bar{B}_s$ mixing is everywhere weaker than the $\rho$-parameter constraint. Likewise, the LEP LFU constraint is everywhere contained within the constraint  from perturbative unitarity.}  
\end{center}
\end{figure}

Our $\ZP$ has large branching ratios for all of these decays channels, for the range of $a$ values that fit the other constraints. However, like other third family $\ZP$ models, this heavy gauge boson has only weak couplings to valence quarks, here governed by the $SU(2)_{\mathrm{global}}^3\times U(1)_X$-breaking spurions, and so its production cross-section in $pp$ collisions is much suppressed (relative to a $\ZP$ boson with family universal couplings). For third family $\ZP$ models with similar relevant couplings to the model we discuss here, the direct searches are all evaded for $M_{\ZP} \gtrsim 2$ TeV or so~\cite{Allanach:2019mfl,Allanach:2020kss}. We leave a precise computation of the direct search bound on the present model for future studies.

The model presented in this paper does, however, differ to previous third family hypercharge models in some crucial aspects, phenomenologically. Firstly, the (necessarily vector-like) coupling $a$ to muons is not fixed by anomaly-freedom, but is an (integer-quantized) parameter of the model (we have shown in \S \ref{sec:UV} how a pair of charged fermions can always be `integrated in' at the TeV scale to restore anomaly cancellation, for any value of $a$). The direct search in the muon channel will put a constraint on the combination of $M_{\ZP}$ and $a$. Moreover, the total width of the $\ZP$ will grow with $a$, both through the partial widths to dimuon and ditauon final states, and requiring the $\ZP$ width be sufficiently narrow will translate to a constraint on the gauge coupling $g_X$ and $a$, which we compute next. Finally, the fact that we gauge an anomalous gauge symmetry means there are novel decays of the $\ZP$ to pairs of SM gauge bosons, which we comment on. The branching ratios for these decays are, however, extremely suppressed.

To summarize, the $Z^\prime$ gauge boson of our $Y_3+a(L_\mu-L_\tau)/6$ model has various decays, which in general can be grouped in 3 categories:
\begin{itemize}
\item Unsuppressed tree-level decays to fermion pairs, $Z^\prime \to f \bar{f}$. These dominate the width of the $\ZP$.
\item Suppressed tree-level decays, where the suppression is due to a factor of the $Z-Z^\prime$ mixing angle, which is proportional to $(M_Z/M_{\ZP})^2 \sim 10^{-4}$. This includes $Z^\prime \to ZH$ and $Z^\prime \to W^+ W^-$.
\item One-loop decays induced by the gauge anomaly. For our model, which features only the quadratic ($\mathcal{A}_{XXY}$) and cubic ($\mathcal{A}_{XXX}$) gauge anomalies, the only kinematically allowed decays of this kind are further suppressed by mixing angles.
Specifically, we have the one-loop decays $Z^\prime \to ZZ$ (which receives contributions from both the $\mathcal{A}_{XXY}$ and the $\mathcal{A}_{XXX}$ anomaly, after $Z-Z^\prime$ mixing) and $Z^\prime \to Z\gamma$ (which receives contributions from $\mathcal{A}_{XXY}$ only, because the $X$ boson does not mix with the photon). In addition to suppression by factors of the $Z-Z^\prime$ mixing angle, as for the previous category, we expect these one-loop decays to be suppressed by a {\em further} factor of $10^{-4}$ at least relative to the $Z^\prime \to f \bar{f}$ channels, as was found in~\cite{Ekstedt:2017tbo}. 
\end{itemize}
The hierarchy of BRs means we only need to consider decays in the first category to get an accurate expression for the $Z^\prime$ width, and thence estimate a rough bound from perturbative unitarity.

The partial width of the $Z^\prime$ into a massless fermion $f_i$ and its antiparticle is given by the formula $\Gamma_{ij} = \frac{C}{24 \pi} |g_{ij}|^2 M_{Z^\prime}$ where the colour factor $C=3$ for quarks and $C=1$ for leptons, in the limit that the $\ZP$ is much heavier than the fermion (which we can safely assume for our TeV scale $\ZP$).
Summing over the SM fermions, we obtain
\be
\frac{\Gamma_{Z^\prime}}{M_{Z^\prime}} = \frac{g_X^2}{144\pi} \left(20 + 4a+a^2 \right).
\ee
The model approaches non-perturbativity when $\Gamma_{Z^\prime}/M_{Z^\prime} \approx 1/3$ or so, when the width of the $Z^\prime$ resonance becomes broad. Thus, perturbativity breaks down for couplings of order
\be
g_X \sim \sqrt{\frac{48 \pi}{20 + 4a+a^2}} 
\ee
We plot this constraint for both $M_{\ZP}=3$ TeV and $M_{\ZP}=6$ TeV, along with the others computed above, in Fig.~\ref{fig:pert}. Both these masses ought to be safe from the LHC direct search bounds. We also fix the quark mixing angle to be $\sin\theta_b = 2|V_{tb} V_{ts}^\ast|/c_d$ in these plots, which makes the $B_s-\bar{B}_s$ mixing and the $\rho$-parameter constraints of comparable strength, allowing one to access the widest range of values for the parameter $a$. 

From the left-hand plot of Fig.~\ref{fig:pert}, for $M_{\ZP}=3$ TeV, there is a reasonably wide range of parameter space for which the couplings remain perturbative, with the parameter $a$ ranging between 6 and 50. At the upper end of this range, for which the $\ZP$ width approaches one third its mass, we find that $v_X:=M_{\ZP}/g_X$ can be pushed as high as 15 TeV or so while satisfying all constraints, as we have used to set our upper limits on both the mass of the charged fermions (\ref{eq:5TeV}) and the scale $\Lambda$ of heavy new physics (\ref{eq:Lambda}).
From the right-hand plot of Fig.~\ref{fig:pert}, we see that a 6 TeV $\ZP$ is just about realisable with perturbative couplings, but the mass cannot be pushed much higher without the $\ZP$ theory becoming strongly coupled. 

\section{Discussion}

In this paper we have considered extending the SM gauge symmetry by $U(1)_X$, where $X$ acts as a linear combination of third family hypercharge and $L_\mu-L_\tau$ on the SM fermions,
$$X=Y_3+\frac{a}{6}(L_\mu-L_\tau).$$ 
Our goal is to develop a simple $U(1)_X$ model for the statistically significant observations of LFUV in $b\to s \ell \ell$ transitions, while (i) preserving the accidental lepton number symmetries of the SM, and (ii) simultaneously shedding light on the flavour problem. 
The $U(1)_X$ charge assignment that we gauge permits only the third family Yukawa couplings, with the renormalisable lagrangian possessing $U(2)^3_\mathrm{global}$ accidental symmetries in the quark sector, while also coupling directly to muons as needed to explain the $b\to s \ell \ell$ anomalies. Any charge assignment with these properties suffers from gauge anomalies, which involve not only $U(1)_X$ but also part of the SM gauge symmetry. In this case, there is a mixed anomaly involving one hypercharge boson and two $X$ bosons, as well as a cubic $U(1)_X$ anomaly. To build a microscopic completion for this model must therefore be done in two stages. 

The first step, which we take in \S \ref{sec:UV}, is to identify suitable chiral matter to restore anomaly cancellation, and this must occur at a scale $\mathcal{O}(10\mathrm{~TeV})$ associated with the $U(1)_X$ breaking (with this scale being fixed by the size of the discrepancies measured in $b\to s \ell\ell$, together with electroweak constraints). We find that a pair of electrically charged fermions, which are chiral under $U(1)_X$, can always be found to do the job. The field content of the model is summarised in Table~\ref{charges}.

A second step must eventually be taken, going deeper into the UV, in order to explain the origins of the light fermion masses and mixings. One way to do this is via vector-like matter at a higher scale $\Lambda = \mathcal{O}(100\mathrm{~TeV})$. In this work we remain agnostic about the dynamics in the UV, and instead analyse the Yukawa sector using an effective field theory of the SM fields and a minimal set of spurions that break the $SU(2)^3_{\mathrm{global}}\times U(1)_X$ symmetries. The quark sector is thus described by a slight variation of models based on $U(2)^3$ breaking that are known to give a good account of quark masses and mixing angles.

On the other hand, one expects a charged lepton Yukawa that is diagonal up to parts per mille corrections, which means the $\ZP$ couplings conserve lepton flavour. Furthermore, using the TeV scale EFT governed by our $U(1)_X$ gauge symmetry, we estimate that the $\Lambda$-scale physics makes only tiny contributions to LFV processes such as $\mu\to e\gamma$ or $\mu \to 3e$, confirming that the SM's accidental lepton number symmetries are well-preserved in the model. This is ultimately a consequence of gauging a component of $L_\mu - L_\tau$ with its tuneable coefficient $a$.

Finally, we perform a phenomenological analysis of the most important constraints on the $\ZP$ boson of our model, which we expect to dominate new physics contributions constrained by experiments. In order to simultaneously satisfy constraints from $B_s-\overline{B_s}$ mixing and the electroweak $\rho$-parameter, the component $a$ of $L_\mu-L_\tau$ must be sufficiently large. Specifically, for a choice of mixing angle between the strange and bottom quarks of $\sin\theta_b\approx 2 |V_{ts}| \approx 0.1$, for which $B_s-\overline{B_s}$ and $\rho$ constraints become roughly equivalent, we require $a \geq 6$, which sets the minimum value of $a$ that is phenomenologically allowed (see Fig.~\ref{fig:const}). For a $\ZP$ mass of 3 TeV, which we expect to be consistent with LHC direct search bounds, we find that $a$ can be as large as $\sim 50$ before the $\ZP$ couplings become non-perturbative (Fig.~\ref{fig:pert}), so the allowed parameter space of the model is wide.

\subsection*{Outlook}

There are many further aspects of the model that need investigating, on both the phenomenological side and the model building side. On the phenomenological side, it remains to accurately compute the bounds on the $\ZP$ from LHC direct searches. We have, in this paper, also ignored the possibility of a coupling $\mathcal{L}\supset -\lambda_{\Phi h} |\Phi|^2|h|^2$ in the scalar potential, which would facilitate a distinctive {\em flavonstrahlung} decay mode~\cite{Allanach:2020kss}.
Furthermore, our phenomenological analysis of the model in this paper has been at the (somewhat crude) level of demonstrating compatibility with various constraints at the 2$\sigma$ level. A more statistically rigorous assessment of the model requires doing a global fit to a combination of electroweak data and $b\to s \ell\ell$ data, as done in {\em e.g.}~\cite{Allanach:2021kzj}. 

An important next step is to study the phenomenology of the heavy chiral fermions $\Psi_{1,2}$, whose masses could naturally lie anywhere between 1 TeV and 30 TeV, and whose existence is mandated by anomaly cancellation. In the version of the model that we have sketched in this paper, these fermions are essentially stable charged particles, which couple only to the neutral gauge bosons $\gamma$, $Z$, and $\ZP$. They will therefore be produced in $pp$ collisions via Drell--Yan production, and would lose energy predominantly via ionization on passing through the detector. At the LHC, existing searches for long-lived particles from ATLAS and CMS are based on a variety of states in certain benchmark SUSY models, of which our colourless states $\Psi_{1,2}$ bear most resemblance to a chargino or (the lightest) slepton, since $\Psi_{1,2}$ do not interact hadronically. Current limits on these particles do not exceed 1 TeV or so~\cite{Aaboud:2019trc} (see also~\cite{Khachatryan:2016sfv,Sirunyan:2020cao}), and it would be interesting to recast these limits for the long-lived particles in our model. Future prospects at the HL-LHC (with ATLAS and CMS searches being complemented also by searches using the MoEDEL detector~\cite{Acharya:2020uwc}) would push these searches further. Even at the upper limit of their mass range, one would expect the charged fermions of this model to be produced at a future collider (FCC-hh) with $\mathcal{O}(100 \mathrm{~TeV})$ centre-of-mass energy.

We also observe that, for the choice $y=-6$, the states $\Psi_{1,2}$ have the same SM quantum numbers as the right-handed electron fields. Given also a SM singlet scalar field $S$ with the appropriate $U(1)_X$ charge, one could then write down a coupling between these heavy fermions and right-handed charged leptons~\cite{Egana-Ugrinovic:2018roi}, facilitating a decay $\Psi_i \to \ell_R S \to \ell_R hh$, where the $S$ decays, say, to two physical Higgs bosons. For the anomaly cancellation story of \S \ref{sec:UV} to work (with integral charges) then requires that $a$ be an odd multiple of 6, corresponding to gauging $X=Y_3+(2n+1)(L_\mu-L_\tau)$, for $n\in\mathbb{Z}$. The experimental bounds computed in this paper would then be consistent with four distinct models, corresponding to $n=0$, 1, 2, or 3. Analysing this subset of our models, for which the now-metastable charged fermions have richer phenomenology, would be another future direction worth exploring. 

In any of these scenarios, it remains to demonstrate that the long-lived charged particles are cosmologically viable, which requires their early-universe abundance be small enough not to spoil the abundances of light nuclei produced during Big-Bang nucleosynthesis (BBN)~\cite{Cyburt:2009pg,Cyburt:2010vz}. Given their natural mass range extends up to 30 TeV, this should be achievable if the reheating temperature after inflation is low enough~\cite{Kudo:2001ie,Drees:2006vh,Drees:2007kk,Takayama:2007du}.

Other challenges and possibilities remain on the model building side. Firstly, one could make the analysis of \S \ref{sec:yuk} more quantitative by setting out an explicit model of the $\Lambda$-scale heavy physics, such as a specific set of $U(1)_X$-charged vector-like fermions. With a UV model in hand, one could calculate the loop-induced contributions to $\Delta F = 2$ processes such as kaon mixing, to properly assess the viability of gauging $X=Y_3+a(L_\mu-L_\tau)/6$ as a model of flavour.

Another feature that requires more detailed study is the origin of neutrino masses within such a model. In Appendix \ref{app:neutrino} we sketch one possible mechanism, which involves three right-handed neutrinos and an additional SM singlet scalar field $\Phi'$. Such a model allows for a sufficiently populated active neutrino mass matrix, but only when the vev of $\Phi'$ is sufficiently big. We have not studied this condition quantitatively in this paper, but introducing the second scalar (with a vev $\langle \Phi' \rangle\approx \langle \Phi \rangle$) will inevitably degrade the lepton number symmetries in our model, which were an attractive feature. The crucial question to settle quantitatively is by `how much' will the neutrino mixings degrade the lepton number symmetries, and whether there is a viable parameter space that adequately fits both the neutrino mixing data and the strict constraints on LFV. 

Finally, incorporating heavy vector-like fermions, as suggested above, would afford another boon. The heavy vector-like leptons, which are needed to explain the electron and muon masses, can also explain (see {\em e.g.}~\cite{Czarnecki:2001pv,Murakami:2001cs,Jegerlehner:2009ry,Kannike:2011ng,Dermisek:2013gta,Allanach:2015gkd,Raby:2017igl,Megias:2017dzd,CarcamoHernandez:2019ydc,Yin:2021yqy}) an observed discrepancy between the muon anomalous magnetic moment and its SM prediction~\cite{Aoyama:2020ynm,Aoyama:2012wk,atoms7010028,Czarnecki:2002nt,Gnendiger:2013pva,Kurz:2014wya,Davier:2017zfy,Keshavarzi:2018mgv,Colangelo:2018mtw,Hoferichter:2019mqg,Davier:2019can,Keshavarzi:2019abf,Melnikov:2003xd,Masjuan:2017tvw,Colangelo:2017fiz,Hoferichter:2018kwz,Gerardin:2019vio,Bijnens:2019ghy,Colangelo:2019uex,Colangelo:2014qya,Blum:2019ugy}. This discrepancy has recently grown in significance with the celebrated Muon $g-2$ measurement announced by  Fermilab in April 2021~\cite{Abi:2021gix}. However, it is important to note that state-of-the-art lattice QCD calculations of $(g-2)_\mu$ are consistent with the Fermilab measurement~\cite{Borsanyi:2020mff}, and so the size of the required new physics contribution remains the subject of ongoing debate. In this context, we save a computation of $(g-2)_\mu$ within our anomalous $\ZP$ model,\footnote{Note that the Green--Schwarz terms discussed in \S \ref{sec:IR} also contribute to $(g-2)_\mu$. But this is a two-loop effect~\cite{Armillis:2008bg} that is suppressed beyond the usual 1-loop $Z^\prime$ contribution, and so will not give a phenomenologically relevant contribution for the heavy $\ZP$ mass of interest to us here.} which requires us specify an explicit heavy sector, for more detailed model building studies in the future.

\subsection*{Acknowledgments}

I am most grateful to Ben Allanach and Marco Nardecchia for many enlightening discussions throughout the course of this project, and for carefully reading previous versions of this manuscript. I also thank Anders Eller Thomsen, Admir Greljo, Nakarin Lohitsiri, Matthew McCullough, David Tong, and members of the Cambridge Pheno Working Group for discussions.
I am supported by the STFC consolidated grants ST/P000681/1 and ST/T000694/1.

\appendix

\section{Anomaly algebra} \label{app:algebra}

In this Appendix we derive the anomaly-free assignment of the BSM chiral fermion representations that was used in \S \ref{sec:UV}. Recall that we wish to cancel the anomalous contributions from the SM fermions given in Eq. (\ref{eq:gauge anomalies}).

We extend the SM matter content by an even number of BSM chiral Weyl fermions, pairs of which have the same hypercharge and are neutral under $SU(3)\times SU(2)_L$, which can therefore be lifted via Yukawa interactions with SM singlet scalars. Let $N$ denote the number of such Weyl pairs, so that there are $2N$ BSM Weyl fermions in total, $N$ of which we take to be left-handed and $N$ of which we take to be right-handed. Let the left-handed Weyl fermions $\psi_i$ have $U(1)_X$ charges of $v_i/6$, $i=1,\dots,N$, and let the right-handed Weyl fermions $\chi_i$ have charges $w_i/6$, $i=1,\dots,N$. At this point we make a simplifying assumption, which is that {\em all} the BSM Weyl fermions have the same hypercharge, which we denote by $y/6$.\footnote{Since the extra fermions are $SU(2)_L$ singlets, their electric charge will also equal $y/6$.} In our chosen normalisation, we look for solutions to the anomaly cancellation equations over the integers. 

Adding together the contributions from the SM fermions and the BSM fermions, anomaly cancellation requires the pair of equations be satisfied,
\begin{align}
6a = y \sum_{i=1}^N \left(w_i^2 - v_i^2 \right), \label{eq:quad_dark} \\
54a = \sum_{i=1}^N \left(w_i^3 - v_i^3 \right). \label{eq:cubic_dark}
\end{align}
The first equation is to cancel the mixed anomaly with hypercharge, and the second to cancel the cubic $U(1)_X$ anomaly.
Together these equations imply, by factorising and eliminating $a$,
\be  \label{eq:dioph}
\sum_{i=1}^N (w_i-v_i) \left[ w_i^2+w_i v_i +v_i^2 - 9y(w_i+v_i) \right] = 0.
\ee
We must take care not to re-introduce a $U(1)_Y^2 \times U(1)_X$ anomaly or a mixed gravitational--$U(1)_X$ anomaly. Because all the $\psi_i$ have the same hypercharge $y$, these two conditions give the single constraint
\be \label{eq:sum}
\sum_i v_i = \sum_i w_i.
\ee
This last condition, together with (\ref{eq:quad_dark}) or (\ref{eq:cubic_dark}), requires the number of BSM fermion pairs $N \geq 2$ for $a \neq 0$, as needed for the essential coupling to muons to be non-zero. 

Continuing, we look for solutions in the simplest case where $N=2$.
Substituting (\ref{eq:sum}) into (\ref{eq:dioph}) to eliminate $v_2$, it turns out that the resulting cubic can be factorised,
\be
(v_1-w_1)(v_1-w_2)\left(18y-3(w_1+w_2)\right)=0.
\ee
The first two roots correspond to trivial solutions in which $y=0$ and hence $a=0$, which simply return the third family hypercharge assignment $X=Y_3$. 
Thankfully, the last root does not force $a=0$, but rather requires only that the condition
\be \label{eq:y_constraint}
y = \frac{1}{6}(w_1+w_2)=\frac{1}{6}(v_1+v_2)
\ee
on the hypercharge quantum number $y$ be satisfied.
Provided $y$ is fixed accordingly, we thus arrive at anomaly-free solutions for {\em any} values of the $U(1)_X$ charges $w_1$, $w_2$, and $v_1$, choosing $v_2 = w_1+w_2 - v_1$. 
For example, take $w_1=1$, $w_2=5$, $v_1=2$, $v_2=4$, meaning that (\ref{eq:y_constraint}) is satisfied for $y=1$. The quadratic anomaly constraint is
$6a = 1^2+5^2-2^4-4^2 = 6$,
while the cubic is
$54a= 1^3+5^3-2^3-4^3 = 54$,
both of which are consistent with $a=1$.

Of course, we instead wish to fix the phenomenologically important parameter $a$ (which recall sets the $U(1)_X$ charge of the muon), and then find a corresponding set of charges $(v_i,w_i)$.
The formula for $a$ is
\be \label{eq:x1}
a = \frac{1}{3}y(w_1-v_1)(w_1+v_1-6y),
\ee
%where $w_1$, $v_1$, and $y$ can be chosen freely. Recall that $a=L_2=E_2$ is the (vector-like) $U(1)_X$ charge of the muon (in units of one sixth), and so this quantity $a$ is a crucial parameter in the phenomenology of the fit to the NCBAs and related constraints.
Now, the difference $w_1-v_1$ that appears as a factor in (\ref{eq:x1}) corresponds to (six times) the $U(1)_X$ charge $X_\Phi$ of a scalar field $\Phi$ which can be used to Higgs all the extra chiral fermions, as we discuss in \S \ref{sec:UV}. Taking the minimal choice $X_\Phi=(w_1-v_1)/6=1/6$, the formula (\ref{eq:x1}) simplifies to $3a=y(1+2v_1-6y)$, which we can invert to find the charges recorded in Eq. (\ref{eq:BSMcharges}) in the main text.
%\begin{align}
%w_1 &= 3y+\frac{3a}{2y}+\frac{1}{2}, \\
%w_2 &= 3y-\frac{3a}{2y}-\frac{1}{2}, \\
%v_1 &= 3y+\frac{3a}{2y}-\frac{1}{2}, \\
%v_2 &= 3y-\frac{3a}{2y}+\frac{1}{2}.
%\end{align}
%One can explicitly check that these charges satisfy the equations (\ref{eq:quad_dark}, \ref{eq:cubic_dark}, \ref{eq:sum}) for anomaly cancellation.
%To ensure that all the charges are integer multiples of one sixth means we must choose a value of $y$ such that $a$ is an odd multiple of $y$,
%\be
%a \in (2\mathbb{Z}+1)y.
%\ee
%For any odd $a$ we can just take $y=1$, while for any even $a$ a suitable value $y=2^k$, $k\in \mathbb{Z}$, can always be found.

\section{One route to neutrino masses} \label{app:neutrino}

In this Appendix we sketch one possible account of neutrino masses within our gauged $X=Y_3+a(L_\mu-L_\tau)/6$ model.
To do so, we extend the field content of Table~\ref{charges} by the additional SM singlet fields written in Table~\ref{charges2}. We include three right-handed neutrinos $\nu_i$, which are SM singlets and charged in the obvious way under $X=Y_3+a(L_\mu-L_\tau)/6$, the addition of which preserves anomaly cancellation. We also introduce an additional SM singlet scalar field $\Phi^\prime$ with $U(1)_X$ charge of $a/6$.

We continue to assume the integer $a \geq 9$ or so, as in \S \ref{sec:LFV} of the main text. By including the second scalar field with charge $a/6$, we can write down both Yukawa and Majorana mass matrices for the neutrinos. Firstly, the Dirac mass matrix for the neutrinos takes the form
\be
M_D \sim \frac{v}{\sqrt{2}}
\left(\begin{array}{ccc}
    \epsilon_\Phi^3 & 0 & 0 \\ 0 & \epsilon_\Phi^3 & 0 \\ \epsilon_{\Phi'} & 0 & 1 \\
  \end{array}\right) 
+ \mathcal{O}(\epsilon_\Phi^3 \epsilon_{\Phi'}, \epsilon_{\Phi'}^2),
\ee
where $\epsilon_{\Phi'} = \langle \Phi^\prime \rangle/\sqrt{2}\Lambda$. The Yukawa matrix for the charged leptons remains the same as in (\ref{eq:lepton_yuk}), only this time the off-diagonal terms we have dropped come in at order $\epsilon_\Phi^3 \epsilon_{\Phi'}$ in our expansion parameters -- thus, for this particular account of neutrino masses, the size of $\epsilon_{\Phi'}$ sets the scale of LFV in the charged lepton sector of the model, and so must be sufficiently small. The other small parameter $\epsilon_\Phi=\mathcal{O}(0.1)$, as in the main text.

\begin{table} 
\begin{centering}
\begin{tabular}{c|c|c|c}
Field & Chirality & $G_{\text{SM}}$ & $6 \times U(1)_X$ \\
\hline
$\nu_1$ & R & $(\mathbf{1},\mathbf{1},0)$ & $0$ \\
$\nu_2$ & R & $(\mathbf{1},\mathbf{1},0)$ & $a$ \\ 
$\nu_3$ & R & $(\mathbf{1},\mathbf{1},0)$ & $-a$ \\
\hline
$\Phi^\prime$ & - & $(\mathbf{1},\mathbf{1},0)$ & $a$
\end{tabular}
\caption{\label{charges2} Additional fields (beyond the content of Table~\ref{charges}) included in a variation of our model that can accommodate massive neutrinos. }
\end{centering}
\end{table}

For the right-handed neutrinos, there are both super-renormalisable dimension-3 Majorana masses, whoses sizes are set by the heavy scale $\Lambda$, as well as dimension-4 mass terms which require an insertion of the second flavon $\Phi^\prime$. The Majorana mass matrix then has the structure
\be
M_R \sim \frac{\Lambda}{\sqrt{2}}
\left(\begin{array}{ccc}
    1 & \epsilon_{\Phi'} & \epsilon_{\Phi'} \\ \epsilon_{\Phi'}  & 0 & 1 \\ \epsilon_{\Phi'} & 1 & 0 \\
  \end{array}\right) 
+ \mathcal{O}(\epsilon_{\Phi'}^2),
\ee
up to coefficients that we expect to be order-one.
Since all the entries in the Majorana mass matrix are much greater than those in the Dirac matrix, we can use the usual see-saw formula to obtain the mass matrix for the light active neutrinos,
\be
M_{\nu_L} \approx -M_D M_R^{-1} M_D^T.
\ee
Noting that $\epsilon_\Phi^3 =\mathcal{O}(10^{-3})\approx y_\mu$, the muon Yukawa, as noted in the main text, we can write
\be
M_{\nu_L} \sim \frac{v^2}{\Lambda}
\left(\begin{array}{ccc}
y_\mu^2 & y_\mu^2\epsilon_{\Phi'} & y_\mu\epsilon_{\Phi'} \\ 
y_\mu^2 \epsilon_{\Phi'}  & y_\mu^2\epsilon_{\Phi'}^2 & y_\mu \\ 
y_\mu \epsilon_{\Phi'} & y_\mu & \epsilon_{\Phi'}^2 \\
  \end{array}\right) 
+ \dots
\ee
up to dimensionless coefficients that we expect to be $\mathcal{O}(1)$, and where we have indicated only the leading order contribution to each matrix element.
We thus expect the largest terms to be the $\left(M_{\nu_L}\right)_{23,32}$ matrix elements, predicting a large neutrino mixing angle $\sin^2\theta_{23}$ consistent with observations. A more detailed study of the lepton sector we have outlined is warranted. For now, we point out that fitting the neutrino mixing data carefully will set a minimum size for $\epsilon_{\Phi'}$ (in order for enough of the matrix elements of $M_{\nu_L}$ to be big enough), which is bounded from above by LFV constraints.

%%%%%%%%%%%%%%%%%%%%%%%%%%%
\bibliographystyle{JHEP-2}
\bibliography{references}
%%%%%%%%%%%%%%%%%%%%%%%%%%%%%%%%%%%%%%%%%%%%%%%%%%%%%%%%%%%%%%%%%%%%%%

\end{document}